\documentclass[a4paper,11pt]{article}
\pdfoutput=1 

\usepackage{jheppub} 

\usepackage[T1]{fontenc} 
\allowdisplaybreaks[2]

\makeatletter
\def\@fpheader{$ $}
\makeatother

\title{\boldmath Scattering Amplitudes and Soft Theorems in Multi-Flavor Galileon Theories}

\author{Karol Kampf}
\author{and Ji\v{r}\'{\i} Novotn\'y}
\affiliation{Institute of Particle and Nuclear Physics, Charles University\\V Hole\v{s}ovi\v{c}k\'{a}ch 2, 180 00 Prague 8, Czech Republic}

\emailAdd{karol.kampf@mff.cuni.cz}
\emailAdd{jiri.novotny@mff.cuni.cz}

\abstract{
In this paper we initiate the study of multi-flavor Galileon theories using the methods of scattering amplitudes. 
We explore this topic from different perspectives and extend the techniques employed so far mainly in the single-flavor case. 
This includes soft theorems, generalized soft theorems with non-trivial right-hand side, Galileon dualities, soft bootstrap and bonus relations. We demonstrate new properties on two examples, the multi-flavor $U(N)$ Galileon and the three-flavor $U(2)/U(1)$ Galileon. 
}

\begin{document} 
\maketitle
\flushbottom
\section{Introduction}
The Galileon-like theories have attracted considerable attention in recent years, mainly due to their connections with various models of modified gravity. Originally the cubic Galileon coupling emerges as the only nontrivial interaction term for the scalar degree of freedom in the decoupling limit of the DGP gravity \cite%
{Dvali:2000hr,Deffayet:2001pu} and similarly in the case of the massive gravity \cite{deRham:2010kj}, where this degree of freedom corresponds to the helicity zero component of the massive graviton in the decoupling limit. 
It was recognized soon \cite{Nicolis:2008in} that the Galileon couplings can be generalized,  preserving at the same time the main features of the cubic one, namely the second-order equations of motion free of the Ostrogradski ghosts and  the invariance of the action with respect to the (first-order) polynomial shift symmetry. 
These features together with the presence of the Vainshtein screening mechanism  \cite{Vainshtein:1972sx} appear to be attractive for cosmological model building and a raising interest followed the seminal paper \cite%
{Nicolis:2008in} where the Galileon Lagrangians have been classified and used for the local large scale modification of the gravity. 
Further generalizations of the original Galileon theories have been developed,  
namely, the extension beyond the flat Minkowski background revealed the connection of the generally covariant Galileon Lagrangians with the scalar-tensor Horndeski theories. 
Moreover, Galileon-like theories with multiple scalars have been discussed \cite{Hinterbichler:2010xn,Padilla:2010de,Padilla:2010ir, Padilla:2012dx, Sivanesan:2013tba, Allys:2016hfl} and classified \cite{Bogers:2018kuw, Bogers:2018zeg} and very recently also in the context of a nonrelativistic quantum field theory \cite{Brauner:2020ezm}.
For a review and further references see e.g. \cite{Curtright:2012gx, Khoury:2013tda}.

From the formal point of view, the Galileon field can be treated as a Nambu-Goldstone boson corresponding to the spontaneous symmetry breaking of the group $Gal(1,D)$ down to the $D$-dimensional Poincare group $ISO(1,D-1)$. 
The first-order polynomial shift symmetry, which is realized nonlinearly, corresponds then to the broken generators of $Gal(1,D)$. 
This nonlinear symmetry has an important consequence on the quantum level, namely the presence of the so-called \emph{enhanced} Adler zero of the (tree-level) scattering amplitudes \cite%
{Cheung:2014dqa,Cheung:2015ota,Cheung:2016drk}. 

The usual Adler zero \cite{Adler:1964um}, i.e. the vanishing of the amplitudes when one of the Goldstone bosons becomes soft, is a well-known feature of the theories with spontaneous symmetry breaking, though it is not an automatic consequence of the Goldstone boson nature of the corresponding particles \cite{Kampf:2019mcd} and under some circumstances can be violated. 
The \emph{enhanced} Adler zero means that the scattering amplitudes behave schematically as $A(p)=O(p^{\sigma})$ for $p\to 0$, where the \emph{soft exponent} $\sigma$ is greater than one. 
Therefore, the amplitude behaves better in the single soft Goldstone boson limit than expected from the very fact of the presence of the Adler zero.
For the general  Galileon we get $\sigma=2$, while by means of an appropriate fine-tuning of the parameters of the Galileon Lagrangian we can construct the \emph{Special Galileon} for which $\sigma=3$ \cite{Cheung:2014dqa,Cachazo:2014xea}. 
In the latter case, there is an additional second-order generalized polynomial shift symmetry of the fine-tuned Lagrangian, which is responsible for the further enhancement of the soft behavior of the scattering amplitudes \cite{Hinterbichler:2015pqa,
Novotny:2016jkh}.

The enhanced soft limit together with \emph{power counting}, which expresses the interrelation of the degree of homogeneity  of the amplitudes in the external momenta with particle multiplicity, can be used as an alternative definition of the (single) Galileon theories. 
The reason is that the Galileon is a member of the set of the \emph{on-shell reconstructible theories}, the tree-level amplitudes of which can be constructed iteratively starting with the lowest-point amplitudes.
As an appropriate tool for this purpose we can use the soft BCFW recurrence relations \cite{Cheung:2015ota}. 
The latter incorporates the general requirements which must be satisfied by any theory, above all the locality and the  one-particle unitarity,  as well as the specific information distinguishing different theories, namely the soft behavior of the amplitudes in the single-particle soft limit. 
The remaining specification of the theory comes from the ansatz for the set of the lowest-point amplitudes which can be fixed by the power counting of the underlying theory.

We can have, therefore, several complementary definitions of the Galileon theories. 
Let us specially mention two of them: the first one is Lagrangian oriented and based on the symmetries of the action, while the second one is amplitude oriented and based on the soft behavior of the amplitudes in the single-particle soft limit. 

In general, the reconstruction of the theory through the soft BCFW recursion is possible only when the theory itself exists. 
Provided such information is missing, a blind application of the recursion starting with some arbitrarily prescribed power counting and soft behavior might lead to inconsistent results that do not correspond to any admissible set of physical scattering amplitudes.
Identification of these inconsistencies is a cornerstone of the method known as the soft bootstrap \cite{Cheung:2016drk,Elvang:2018dco,Elvang:2020lue}, which has been used as a tool for probing the landscape of the effective field theories for the Goldstone bosons.
While the case of single-flavor Goldstone boson theories has been completely classified using this method in \cite{Cheung:2016drk}, an exhaustive analogous classification of the multi-flavor case is still missing.

As far as the multi-flavor Galileon-like theories are concerned, there exists a nice classification based on symmetry considerations \cite{Bogers:2018kuw,Bogers:2018zeg}. 
The theories are classified there according to the possible extension of the $Gal(1,D)$ group with additional generators and according to the scheme of their spontaneous breaking.
Using the coset construction, the Lagrangians are obtained and their nonlinear symmetries are identified.  
However, the complementary amplitude-oriented bottom-up analysis based exclusively on the soft theorems has not been considered yet.
The aim of this paper is to partially fill this gap and  discuss the general properties of the multi-flavor Galileons from this point of view. 

Namely, we will show that unlike the single-flavor case, the interrelation between the symmetries and the form of the soft theorems is not straightforward and that the enhanced Adler zero can be violated under some circumstances. 
As a particular example we will introduce a theory satisfying the soft theorem  which  is an analogue of that for the $U(1)$ fibrated $CP^{N-1}$ sigma model \cite{Kampf:2019mcd}, now for the Galileon power counting.
As another example we will discuss the case of $U(N)$ Galileon, the amplitudes of which allow for stripping the flavor structure and formulating the nontrivial soft theorem directly for the stripped amplitudes.
We will also develop a modification of the soft bootstrap method for Galileon-like theories based on the bonus relation.
This approach allows us to provide a complete classification of the multi-flavor Galileon theories with an enhanced soft limit in the case of two flavors.
We also formulate a conjecture concerning the uniqueness of such theories for a general number of flavors and give numerical evidence for its validity.
We also classify possible multi-flavor analogues of the Special Galileon.

The paper is organized as follows. 
In Section~\ref{sec:2} we introduce the multi-flavor Galileons and discuss their basic properties. 
In Section~\ref{sec:3} we give an elementary proof of the soft theorem for general multi-Galileon. 
In Section~\ref{section_duality} we discuss the possible duality transformations in the multi-Galileon theories with a particular focus on the validity of the soft theorem. 
In Section~\ref{sec:5} we discuss the on-shell reconstruction of the multi-Galileon amplitudes and introduce the bonus relations which can be used for probing the landscape of the possible multi-Galileon theories. 
Section~\ref{applications_section} is devoted to the partial classification of the two-flavor case and of the multi-flavor Special Galileons. 
In Section~\ref{sec:7} we give a survey of the numerical analysis of the multi-Galileons up to five flavors and of the special case of an additional $U(N)$ symmetry. 
In Section~\ref{sec:summary} we summarize the results. 
Some technical issues are postponed to Appendices~\ref{kinematical factors}, \ref{number_2nd_type_terms} and \ref{appendix_bonus_relations}.

\section{General multi-Galileon theory}\label{sec:2}
 
In this section we concentrate on the Lagrangian-oriented definition of the multi-flavor Galileon theories. We introduce a Lagrangian which obeys the requirement of the invariance of the action with respect to the multi-flavor version of the single-Galileon polynomial shift symmetry and which is a straightforward generalization of the single Galileon case.
In comparison to the classification \cite{Bogers:2018kuw,Bogers:2018zeg}, we will neither assume any additional internal symmetry acting on the multi-Galileon multiplet, nor the possible additional broken generators on top of those responsible for the polynomial shift.

\subsection{Lagrangian and symmetries}

Let us remind first the single-Galileon Lagrangian in $D$ dimensions, which is, in general, a linear combination of $D+1$ terms with an increasing number of fields (up to $D+1$).
Since we are interested in the well-behaved quantum version of the theory, we will omit the tadpole and fix the canonical normalization of the kinetic term.
Then the most general form of the Lagrangian reads
\begin{equation}
\mathcal{L} =\frac{1}{2}\partial \phi\cdot \partial \phi 
+\sum_{n=3}^{D+1}\frac{\left( -1\right) ^{D+n}}{n!\left( D-n+1\right) !}%
\,\lambda _{n}\,\varepsilon ^{\mu _{1}\ldots \mu
_{D}}\varepsilon ^{\nu _{1}\ldots \nu _{D}}\phi
\prod\limits_{j=1}^{n-1}\partial _{\mu _{j}}\partial _{\nu _{j}}\phi
\prod\limits_{k=n}^{D}\eta _{\mu _{k}\nu _{k}}\,,  \label{Lagrangian_single}
\end{equation}
where $\lambda_n$ are the coupling constants and the combinatoric factor is introduced for further convenience. The  Lagrangian (\ref{Lagrangian_single}) is invariant (up to a total derivative) with respect to the polynomial shift symmetry 
\begin{equation}
    \phi\rightarrow\phi+a+b\cdot x\,.
    \label{polynomial_shift}
\end{equation}
Though we can use integration by parts and rewrite the above Lagrangian into various different forms, namely this one is well suited for the multi-flavor generalization. 
Let us therefore write the most general Lagrangian of the $N-$Galileon theory in $D$
dimensions in the form\footnote{%
Here and in what follows, the sum over repeated flavor indices is tacitly
assumed.}%
\begin{eqnarray}
\mathcal{L}_{N} 
 &=&\frac{1}{2}\partial \phi _{i}\cdot \partial \phi _{i} 
\notag \\
&&+\sum_{n=3}^{D+1}\frac{\left( -1\right) ^{D+n}}{n!\left( D-n+1\right) !}%
\lambda _{i_{1}\ldots i_{n}}\varepsilon ^{\mu _{1}\ldots \mu
_{D}}\varepsilon ^{\nu _{1}\ldots \nu _{D}}\phi
_{i_{n}}\prod\limits_{j=1}^{n-1}\partial _{\mu _{j}}\partial _{\nu _{j}}\phi
_{i_{j}}\prod\limits_{k=n}^{D}\eta _{\mu _{k}\nu _{k}},
\label{Lagrangian_basic}
\end{eqnarray}%
which was suggested for the first time in \cite{Padilla:2010de}.
In this formula, 
$
\lambda _{i_{1}\ldots i_{n}}=\lambda _{i_{\sigma \left( 1\right) }\ldots
i_{\sigma \left( n\right) }}
$
is totally symmetric rank-$n$ tensor in the flavor space.
The Lagrangian $\mathcal{L}_{N}$ is invariant with respect to the multi-Galileon transformation%
\begin{equation}
\phi _{j}\rightarrow \phi _{j}+a_j+ b_{j}\cdot x.  \label{multi_Galileon_symmetry}
\end{equation}%
Note that exactly this general form of the multi-flavor Lagrangian has also been found as the result of the algebraic classification performed in \cite{Bogers:2018kuw,Bogers:2018zeg} where it corresponds to the generalized non-twisted Galileon theory. 
However, here we do not assume the totally symmetric tensors $\lambda _{i_{1}\ldots i_{n}}$ to be invariant with respect to some representation of the additional internal symmetry group and take them completely unconstrained.

Let us  note that the single-flavor  Lagrangian  (\ref{Lagrangian_single}) is equivalent up to the surface terms to the following one
\begin{eqnarray}
\widetilde{\mathcal{L}} 
&=&\frac{1}{2}\partial \phi\cdot \partial \phi +\sum_{n=3}^{D+1}%
\frac{\left( -1\right) ^{n}}{2\left( n-1\right) !}\lambda _{n}\left( \partial \phi\cdot \partial \phi \right)
\delta _{\nu _{1}\ldots \nu _{n-2}}^{\mu _{1}\ldots \mu
_{n-2}}\prod\limits_{j=1}^{n-2}\partial _{\mu _{j}}\partial ^{\nu _{j}}\phi\,,
\label{Lagrangian_single_alternative}
\end{eqnarray}
where 
\begin{equation}
\delta _{\nu _{1}\ldots \nu _{n}}^{\mu _{1}\ldots \mu _{n}}=\sum_{\sigma \in
S_{n}}\mathrm{sign}\,\sigma ~\delta _{\nu _{\sigma \left( 1\right) }\ldots \nu
_{\sigma \left( n\right) }}^{\mu _{1}\ldots \mu _{n}}
\end{equation}%
is the generalized Kronecker delta. 
This form might suggest that there is an alternative possible multi-flavor generalization with the Lagrangian
\begin{equation}
\widetilde{\mathcal{L}}_{N}=\frac{1}{2}\partial \phi _{i}\cdot \partial \phi
_{i}+\sum_{n=3}^{D+1}\frac{\left( -1\right) ^{n}}{2\left( n-1\right) !}%
\widetilde{\lambda }_{i_{1}i_{2},i_{3}\ldots i_{n}}\left( \partial \phi
_{i_1}\cdot \partial \phi _{i_2}\right) \delta _{\nu _{3}\ldots \nu
_{n}}^{\mu _{3}\ldots \mu _{n}}\prod\limits_{j=3}^{n}\partial _{\mu
_{j}}\partial ^{\nu _{j}}\phi _{i_{j}}
\label{Lagrangian_basic_alternative}
\end{equation}%
with $\widetilde{\lambda }_{i_{1}i_{2},i_{3}\ldots i_{n}}$ symmetric separately
with respect to the first two indices and the remaining $n-2$ indices. 
However, one can easily show that such a multi-flavor generalization would not be symmetric with respect to (\ref{multi_Galileon_symmetry}).
Since we take the polynomial shift symmetry (\ref{multi_Galileon_symmetry}) as the defining property for the multi-flavor Galileon theories in this section, we will omit this possibility and \emph{define}  the multi-flavor Galileons by the Lagrangian (\ref{Lagrangian_basic}). 

Note, however, that for $\widetilde{\lambda }_{i_{1}i_{2},i_{3}\ldots i_{n}}=\lambda _{i_{1}i_{2}i_{3}\ldots i_{n}}$ totally symmetric, the Lagrangians (\ref{Lagrangian_basic_alternative}) and (\ref{Lagrangian_basic}) are equivalent up to the surface terms.

\subsection{The Feynman rules}

The momentum-space Feynman rules for the $n$-point vertices derived from the Lagrangian (\ref{Lagrangian_basic}) are\footnote{Here and in what follows, the hat symbol indicates the omission of the corresponding momentum from the set.}
\begin{equation}
V_{n}\left( p_{1},\ldots ,p_{n}\right) _{i_{1}\ldots i_{n}}=\lambda
_{i_{1}\ldots i_{n}}G\left( p_{1},\ldots ,\widehat{p_{i}},\ldots p_{n}\right)\,,
\label{vertex}
\end{equation}
where 
\begin{equation*}
G\left( p_{1},\ldots ,\widehat{p_{i}},\ldots p_{n}\right) =\det \left(
p_{k}\cdot p_{l}\right) _{k,l\neq i}
\end{equation*}
is a Gram determinant of arbitrary $n-1$ (different) momenta from the set $\{ p_{1},\ldots ,p_{n}\} $ (obtained by removing the $i$-th line
and $i$-th column form the $n\times n$ Gram determinant $G(p_{1},\ldots ,p_{n})$, where $i$ is arbitrary). Note the factorization of the flavor and kinematic factors, which is a result of total symmetry of both $\lambda _{i_{1}\ldots i_{n}}$ and $G(p_{1},\ldots ,\widehat{p_{i}},\ldots p_{n}) $ with respect to the
permutations.
This allows us to relate simply the properties of the Feynman graphs of the multi-flavor Galileons with those of the single-flavor case.

Indeed, the corresponding Feynman rules in the single-Galileon  case (\ref{Lagrangian_single}) are similarly
\begin{equation}
V_{n}\left( p_{1},\ldots ,p_{n}\right) =\lambda _{n}G\left( p_{1},\ldots ,\widehat{p_{i}},\ldots p_{n}\right)\,.
\end{equation}
We see, that the structure of the contribution $A_{\Gamma }(1^{i_{1}},2^{i_{2}},\ldots ,n^{i_{n}})$ of any graph $\Gamma $ in the 
$N-$Galileon theory into the scattering amplitudes $A_n(1^{i_{1}},2^{i_{2}},\ldots ,n^{i_{n}})$ is factorized as (here and in what follows we often use the condensed notation for the particle momenta and flavors $(p_i,j_i)\rightarrow i^{j_i}$)
\begin{equation}
A_{\Gamma }(1^{i_{1}},2^{i_{2}},\ldots ,n^{i_{n}})=\lambda _{\Gamma }\left(
i_{1},\ldots ,i_{n}\right) A_{\Gamma }(1,2,\ldots ,n)\,.
\end{equation}
Here $A_{\Gamma }(1,2,\ldots ,n)$ is the contribution of the graph with the
same topology as $\Gamma $ (with the same momenta assignment to the external
legs, but without flavor indices) in $N=1$ theory with all $\lambda
_{n}\rightarrow 1$, and $\lambda _{\Gamma }\left( i_{1},\ldots ,i_{n}\right) 
$ is the corresponding ``graph in the flavor space'' with ``vertices'' $\lambda
_{i_{1}\ldots i_{n}}$ and ``propagators'' $\delta _{ij}$ and the same topology
as $\Gamma $ with the same flavor indices assigned to the external legs (but
now without momenta). 

Therefore, any property which can be proved for
individual Feynman graphs in single Galileon theory is also valid in the
multi-Galileon case. But the properties for which it is necessary to have
cancellation between different graphs within the single Galileon theory
cannot be straightforwardly generalized to the multi-Galileon case.
Nevertheless, the ``partial amplitudes'' $A_{\Gamma }(1,\ldots, n)$ form a
kinematical basis for all the $N-$Galileon amplitudes.

\section{Soft theorem for multi-flavor Galileon theories}\label{sec:3}
 
The Lagrangian for the multi-flavor Galileon introduced in the previous section corresponds to the effective theory of the Goldstone bosons associated with the spontaneous symmetry breaking of the polynomial shift symmetry (\ref{multi_Galileon_symmetry}).
In fact, in a straightforward analogy with the single-flavor case, the individual terms of the Lagrangian represent the Wess-Zumino terms constructed from the components of the Maurer-Cartan form after imposing the Inverse Higgs Constraint in order to eliminate the redundant fields associated with the generators of the linear terms in the transformation (\ref{multi_Galileon_symmetry}).
Therefore we expect a presence of the (possibly enhanced) Adler zero for the scattering amplitudes.
In the case of single-Galileon the polynomial shift symmetry (\ref{polynomial_shift}) is responsible for the enhanced Adler zero with the soft exponent  $\sigma=2$. 
However, for the multi-flavor case such a correspondence is not straightforward.
The reason is that in the single Galileon case the enhanced soft behavior is generally not a property valid for individual Feynman graphs but it is rather a result of delicate cancellations between different Feynman graphs contributing to the amplitude.
As we mentioned in the previous section, namely such properties cannot be simply generalized to the multi-flavor case.

The situation when the polynomial shift symmetry does not automatically imply the (possibly enhanced) Adler zero for the Goldstone boson amplitudes is not entirely new. 
For instance, quite recently it was found that the Adler zero is violated for the $U(1)$-fibrated $CP^{N-1}$ nonlinear sigma model \cite{Kampf:2019mcd, preparated}.
Nevertheless, the symmetry is in such a case still powerful enough to ensure validity of a new type of a soft theorem for the Goldstone bosons scattering amplitudes, schematically
\begin{eqnarray}
\lim_{p\rightarrow 0}A_{n+1}(p^{i},1^{i_1},\dots,n^{i_n})=\sum_{j=1}^{n}\sum_{k=1}^{N}C^{i}_{i_j,k}A_{n}(1^{i_1},\dots,j^{k},\dots,n^{i_n})\,,
\end{eqnarray}
where $C^{i}_{j,k}=-C^{i}_{k,j}$ is a set of momentum independent constants.
Note that then the right-hand side is compatible with power counting of the nonlinear sigma model.
In the general case, the power counting  is usually fixed by the parameter 
\begin{equation}
 \rho\equiv\frac{d-2}{n-2}\,, 
 \label{rho_parameter}
\end{equation}
where $d$ is the degree of homogeneity of the amplitudes as a function of the external momenta and $n$ is the particle multiplicity. 
For the nonlinear sigma models $\rho=0$.

Then there is a natural question, whether there exists a theory of Goldstone bosons satisfying some  generalization of this  soft theorem, say that of the schematic form
\begin{eqnarray}
A_{n+1}(p^{i},1^{i_1},\dots,n^{i_n})\overset{p\rightarrow 0}{=}\sum_{j=1}^{n}\sum_{k=1}^{N}(p\cdot p_{j})^{L}D^{i}_{i_j,k}A_{n}(1^{i_1},\dots,j^{k},\dots,n^{i_n})
\label{generalized_soft_theorem}
\end{eqnarray}
for some positive $L$ and for some set of constants $D^{i}_{j,k}$ (not necessarily antisymmetric in $(j,k)$). 
Clearly, provided all the amplitudes of the theory are characterized by the same constant power-counting parameter $\rho$ defined by (\ref{rho_parameter}), the compatibility of (\ref{generalized_soft_theorem}) requires $\rho=2L$.

Since for the Galileons we have $\rho=2$, the natural candidate for the nontrivial soft theorem for multi-flavor Galileons is (\ref{generalized_soft_theorem}) with $L=1$.
In this section we will prove that this conjecture is in fact  correct and that (\ref{generalized_soft_theorem}) with $L=1$  is realized by multi-flavor Galileons with those Lagrangians (\ref{Lagrangian_basic}) that have nonvanishing cubic couplings.

\subsection{Generalization of the soft theorem\label{soft_theorem_section}}

Let us be a little bit less schematic here and give the formula (\ref{generalized_soft_theorem}) more precise meaning.
To preserve the momentum conservation at each step, we have to specify more carefully the way we take the single-particle soft limit.
This can be done as follows.
Let $p_{i}$, $i=1,\ldots ,n$ is a fixed $n$-particle configuration satisfying
\begin{equation}
p_{i}^{2}=0\,,\qquad\sum_{i=1}^{n}p_{i}=0  \label{momentum conservation}
\end{equation}
and $p$ be an arbitrary on shell momentum, $p^2=0$. 
The soft limit of $\left( n+1\right)$%
-point amplitude is then realized  as the $t\rightarrow 0$ limit of the one-parametric
deformation of the momenta
\begin{equation}
p \rightarrow tp\,,\qquad
p_{i} \rightarrow p_{i}(t)\,,
\end{equation}
satisfying the boundary condition $p_{i}(0) =p_{i}$, 
and preserving  the  $(n+1)$-point momentum conservation
\begin{equation}
tp+\sum_{i=1}^{n}p_{i}(t) =0.
\end{equation}%
Provided the deformed momenta are kept on shell for each value of $t$, i.e.
\begin{equation}
    p^{2} =p_{i}^{2}(t)=0,
\end{equation}
then the deformed amplitude
$A_{n+1}(tp^{i},{1}(t)^{i_1},\dots,{n}(t)^{i_n})$ 
is well defined for all $t$
and the soft theorem (\ref{generalized_soft_theorem}) can be reformulated for $L=1$ as
\begin{eqnarray}
\lim_{t\to 0}A_{n+1}(tp^{i},{1}(t)^{i_1},\dots,{n}(t)^{i_n})&=&0,\notag\\
\lim_{t\to 0}\frac{{\mathrm d}}{{\mathrm d}t}A_{n+1}(tp^{i},{1}(t)^{i_1},\dots,{n}(t)^{i_n})&=&\sum_{j=1}^{n}\sum_{k=1}^{N}(p\cdot p_{j})D^{i}_{i_j,k}A_{n}(1^{i_1},\dots,j^{k},\dots,n^{i_n}).
\label{soft_theorem_multi_G}
\end{eqnarray}

Let us now proceed with the proof of such a soft theorem for the multi-flavor Galileon theories.
Assume a general graph $\Gamma$ which contributes to the amplitude $A_{n+1}$ 
and suppose that the soft external momentum $tp$ is attached to the vertex 
$V_{k}(tp,q_{1},\dots,q_{k-1})_{ij_{1}\dots j_{k-1}}$, where 
$q_1,\dots,q_{k-1}$ 
are the remaining external or internal hard momenta entering the vertex.
Since the multi-Galieon vertices (\ref{vertex}) are
quadratic in (all) the  momenta, we get 
\begin{equation}
V_{k}(tp,q_{1},\dots,q_{k-1})_{ij_{1}\dots j_{k-1}}=O(t^{2})\,,
\end{equation}%
irrespectively of the nature of the hard momenta $q_1,\dots,q_{k-1}$.
For $k>3$ or when all the hard momenta $q_1,\dots,q_{k-1}$ 
are  internal, all the propagators of the graph are regular when 
$t\to 0$ 
and as a consequence, the contribution of the graph $\Gamma$ 
behaves as 
$O(t^2)$ 
in the soft limit.
Therefore, in theory without cubic vertices we have $O(t^{2})$
behavior in the single soft limit graph by graph.

\begin{figure}[tb]
  \centering
    \includegraphics[scale=0.75]{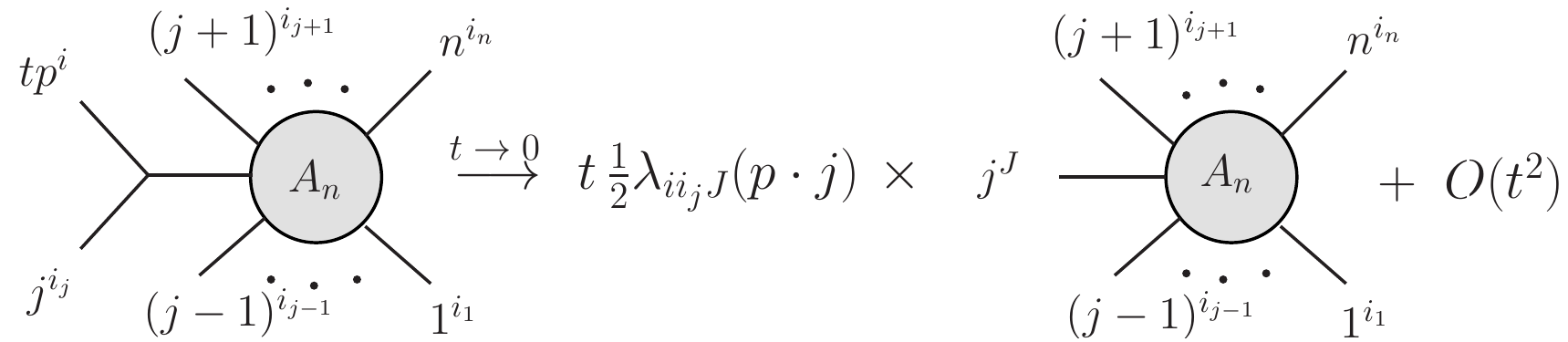}
    \caption{The graphs with a nontrivial single-particle soft limit}
    \label{fig:soft}
\end{figure}

When the cubic vertices are present, we can conclude that the only
nontrivial contribution to the single soft limit of the $\left( n+1\right)$%
-point amplitude 
$
A_{n+1}\left( tp^{i},1(t)^{i_{1}},2(t)^{i_{2}},\ldots ,n(t)^{i_{n}}\right)
$
comes from the graphs when the soft particle $(tp,i)$ is attached to the $j$-th external line of the $n$-point semi-on-shell amplitude 
$A_{n}\left( 1(t)^{i_{1}},\ldots,(j(t)+tp)^{J},\ldots ,n(t)^{i_{n}}\right)$ by three-point vertex (see Fig. \ref{fig:soft}).
The corresponding contribution is
\begin{eqnarray}
&&\mathrm{i}V_{3}\left( tp,j(t),-(j(t)+tp)\right)_{ii_jJ} \frac{1}{(j(t)+tp)^{2}} \mathrm{i}A_{n}\left( 1(t)^{i_{1}},\ldots
,(j(t)+tp)^{J},\ldots ,n(t)^{i_{n}}\right)\notag \\
&&=\frac{1}{2}\lambda _{ii_{j}J}\left( tp\cdot j(t)\right)A_{n}\left( 1(t)^{i_{1}},\ldots
,(j(t)+tp)^{J},\ldots ,n(t)^{i_{n}}\right).
\end{eqnarray}
Summing all such contributions we get finally
\begin{eqnarray}
&&A_{n+1}\left( tp^{i},1(t)^{i_{1}},2(t)^{i_{2}},\ldots ,n(t)^{i_{n}}\right) \label{soft_theorem_galileon}\\
&&\overset{t\rightarrow 0}{=}t\sum_{j=1}^{n}\frac{1}{2}\lambda
_{ii_{j}J}(p\cdot j) A_{n}\left( 1^{i_{1}},2^{i_{2}},\ldots
,(j-1)^{i_{j-1}},j^{J},(j+1)^{i_{j+1}},\ldots ,n^{i_{n}}\right) +O\left(
t^{2}\right)\,.\notag
\end{eqnarray}
The multi-flavor Galileon with the Lagrangian (\ref{Lagrangian_basic}) represents therefore a realization of a theory with the soft theorem (\ref{soft_theorem_multi_G}) where now $D^{i}_{j,k}=\frac{1}{2}\lambda_{ijk}$.
Note that in the single-flavor case this formula simplifies to
\begin{eqnarray}
&A_{n+1}\left( tp,1(t),2(t),\ldots ,n(t)\right) 
\overset{t\rightarrow 0}{=}t\frac{1}{2}\lambda
_{3}A_{n}\left( 1,2,\ldots
 ,n\right)\sum_{j=1}^{n}(p\cdot j)  +O\left(
t^{2}\right),
\end{eqnarray} 
and finally
\begin{equation}
A_{n+1}\left( tp,1(t),2(t),\ldots ,n(t)\right) 
\overset{t\rightarrow 0}{=} O\left(
t^{2}\right),
\end{equation}
as a consequence of the momentum conservation.
Provided $\lambda_3\ne 0$, the usual soft theorem for the single Galileon is a result of the cancellation between different Feynman graphs, as we have mentioned above.

\subsection{Examples of  multi-flavor Galileon theories and their soft theorems}\label{sec:explicitexamples}

In this subsection we present two explicit examples of multi-flavor Galileon theories and the corresponding soft theorems. First, we will discuss the $U(N)$ Galileon, which has an additional internal symmetry on top of the multi-flavor polynomial shift symmetry. The second example is a simple three-flavor model, which we call $U(2)/U(1)$ Galileon, with nontrivial soft theorem, which is a $\rho=2$ analog of the one for the  fibrated $CP^1$ nonlinear sigma model established in \cite{Kampf:2019mcd}.

\subsubsection{The $U(N)$ Galileon}

The $U(N)$ Galileon was introduced in \cite{Padilla:2010ir, Cheung:2016drk} and
its Lagrangian, which is explicitly invariant with respect to the $U(N)$ transformations, can be written in the form\footnote{As discussed in \cite{Padilla:2010ir, Cheung:2016drk}, a more general form of the Lagrangian  with multiple traces is possible without spoiling the $U(N)$ invariance. Here we restrict ourselves to the single trace terms in order to ensure the possibility of stripping off the flavor structure form the amplitudes; see below for details.} (we use the notation $%
\langle \cdot \rangle \equiv \mathrm{Tr}\left( \cdot \right) $ to simplify
the formulas in what follows)
\begin{equation}
\mathcal{L}_{U(N)}=\frac{1}{2}\langle \partial \phi \cdot \partial \phi \rangle
+\sum_{n=3}^{D+1}\frac{\left( -1\right) ^{n-1}}{n}\lambda _{n}\delta _{\nu
_{1}\ldots \nu _{n-1}}^{\mu _{1}\ldots \mu _{n-1}}\langle \phi\,\partial
_{\mu _{1}}\partial ^{\nu _{1}}\phi \ldots \partial _{\mu _{n-1}}\partial
^{\nu _{n-1}}\phi \rangle\,.  \label{SU(N)_Galileon_Lagrangian}
\end{equation}
Here $\phi =\sum_{a=0}^{N^2-1}\phi _{a}T^{a}$, where $T^{a}$, $a=0,1,\dots,N^2-1$ are the Hermitian generators of 
$U(N)$ in the defining representation, normalized as
\begin{equation}
\langle T^{a}T^{b}\rangle =\delta ^{ab},
\end{equation}
and the $U(N)$ symmetry is realized linearly according to
\begin{equation}
\phi \rightarrow U\phi U^\dagger\,,\quad U\in U(N)\,.
\end{equation}
Note that the Lagrangian (\ref{SU(N)_Galileon_Lagrangian}) can be rewritten
in the form (\ref{Lagrangian_basic}) with the totally symmetric couplings
\begin{equation}
\lambda _{a_{1}\ldots a_{n}}=\frac{\lambda _{n}}{n}\sum_{\sigma \in
S_{n}}\langle T^{a_{\sigma \left( 1\right) }}\ldots T^{a_{\sigma \left(
n\right) }}\rangle .  \label{SU(N)_n_point_vertex}
\end{equation}
For the three-point coupling we get especially
\begin{equation}
\lambda _{abc}=\lambda _{3}\bigl\langle T^{a}\{ T^{b},T^{c}\} \bigr\rangle =
\sqrt{2}\lambda _{3}d^{abc}.  \label{SU(N)_3pt_vertex}
\end{equation}
The Feynman rules for the interaction vertices read then (cf. (\ref{vertex}))
\begin{equation}
V_{n}\left( 1,2,\ldots ,n\right) _{a_{1}\ldots a_{n}}=\frac{\lambda _{n}}{n}
\sum_{\sigma \in S_{n}}\langle T^{a_{\sigma \left( 1\right) }}\ldots
T^{a_{\sigma \left( n\right) }}\rangle G\left( 1,\ldots ,\widehat{i},\ldots
,n\right)\,,
\end{equation}
and using the completeness relation for the $U(N)$ generators
\begin{equation}
\sum_{a=1}^{N^{2}}\langle XT^{a}\rangle \langle T^{a}Y\rangle =\langle
XY\rangle\,,
\end{equation}
the usual stripping of the tree-level amplitudes can be proved, namely 
\begin{equation}
A_{n}\left( 1^{a_{1}},\ldots ,n^{a_{n}}\right) =\sum_{\sigma \in S_{n}/%
\mathbb{Z}_{n}}\langle T^{a_{\sigma \left( 1\right) }}\ldots T^{a_{\sigma \left(
n\right) }}\rangle \mathcal{A}_{n}\left( \sigma \left( 1\right) ,\ldots
,\sigma \left( n\right) \right)\,.
\label{stripped_amplitudes_definition}
\end{equation}
Here the stripped amplitudes $\mathcal{A}_{n}\left( 1,\ldots ,n\right) $
have cyclic symmetry and can be constructed using the cyclically ordered
Feynman rules with vertices
\begin{equation}
V_{n}\left( 1,2,\ldots ,n\right) _{a_{1}\ldots a_{n}}=\lambda _{n}G\left(
1,\ldots ,\widehat{i},\ldots ,n\right) .
\end{equation}
Therefore, the graphs which contribute to the stripped amplitudes can be
identified with the cyclically ordered graphs in the single-Galileon theory
with couplings $\lambda _{n}$. This fact can be used to derive the soft
theorem for the stripped amplitudes in the same way as in Section \ref{soft_theorem_section}.  
The
only exception is that, due to the cyclic ordering, there are only two types
of graphs with propagators singular in the soft limit, namely those with two
particle poles $(p+p_{1})^{2}\rightarrow 0$ and $(p+p_{n})^{2}\rightarrow 0$.
Only these can contribute to the right-hand side of the soft theorem. As a
result we get
\begin{equation}
\mathcal{A}_{n+1}\left( tp,1(t) ,\ldots ,n(t)
\right) \overset{t\rightarrow 0}{=}t\frac{\lambda _{3}}{2}\left[ (p\cdot
n)+(p\cdot 1)\right] \mathcal{A}_{n}\left( 1,\ldots ,n\right) +O\left(
t^{2}\right) .
\label{eq:softstrip}
\end{equation}
This soft theorem for the stripped amplitudes $\mathcal{A}_{n}\left(
1,\ldots ,n\right) $ is universal, however, for the full permutationally
invariant amplitudes $A_{n}\left( 1^{a_{1}},\ldots ,n^{a_{n}}\right) $ the
concrete form of the soft theorem depends on $N$.

\subsubsection{$U(2)/U(1)$ Galileon}\label{subsection:u2u1gal}

In this subsection we present a simple example of the three-flavor Galileon
Lagrangian with a non-trivial soft theorem, which mimics the soft theorem of
the fibrated $CP^{1}$ nonlinear sigma model introduced in \cite{Kampf:2019mcd}. On the top of the
polynomial shift symmetry, this model also has additional $U(1)$ symmetry responsible for the charge conservation. 
Let us define the following Hermitian matrix\footnote{Here and in what follows $\sigma _{i}$, $i=1,2,3$ are the Pauli matrices.} 
\begin{equation}
\phi =\left( 
\begin{array}{cc}
\phi _{0} & \sqrt{2}\phi _{-} \\ 
\sqrt{2}\phi _{+} & \phi _{0}
\end{array}
\right) =\phi _{0}\mathbf{1}+\phi _{1}\sigma _{1}+\phi _{2}\sigma _{2}\,,
\end{equation}
where $\phi _{0}$ and $\phi _{\pm }=\left( \phi _{1}\pm \mathrm{i}\phi
_{2}\right) /\sqrt{2}$ are neutral and charged scalar fields respectively.
The Lagrangian of our model is then
\begin{equation}
\mathcal{L}_{U(2)/U(1)}=\frac{1}{4}\langle \partial \phi \cdot \partial \phi \rangle
+\frac{1}{4}\sum_{n=3}^{D+1}\frac{\left( -1\right) ^{n-1}}{n}\lambda
_{n}\delta _{\nu _{1}\ldots \nu _{n-1}}^{\mu _{1}\ldots \mu _{n-1}}\langle
\phi\, \partial _{\mu _{1}}\partial ^{\nu _{1}}\phi \ldots \partial _{\mu
_{n-1}}\partial ^{\nu _{n-1}}\phi \rangle\,,
\end{equation}
and fits the general form for the three-flavor Galileon (\ref{Lagrangian_basic}) with special
values of couplings $\lambda _{a_{1}\ldots a_{n}}$. 
On contrary to the
previous case of $U(N)$ Galileon, the matrices $(\mathbf{1}, \sigma_{1}, \sigma_{2})$ are not closed under commutation, but
rather under graded commutators. 
Indeed, defining
\begin{equation}
\sigma _{\pm }=\frac{1}{2}\left( \sigma _{1}\pm \mathrm{i}\sigma _{2}\right), 
\end{equation}
the generators form the $SL(1|1)$ superalgebra
\begin{equation}
\left\{ \sigma _{+},\sigma _{-}\right\} =\mathbf{1},~~\left[ \sigma _{\pm },\mathbf{1}\right] =\left[\mathbf{1},\mathbf{1}\right] =0,~~\left\{ \sigma _{\pm },\sigma _{\pm
}\right\} =0.
\end{equation}
Note also that we can interpret the matrix $u(\phi) = \exp( \mathrm{i}\phi)$ as a parametrization of the coset $U(2)/U(1)$ 
where $U(1)$ is generated by $\sigma_{3}$ and $U(2)$ is spanned by $(\mathbf{1},\,\sigma_{i})$.
The $U(1)$ symmetry with respect to the transformation
\begin{equation}
\phi \rightarrow U(\alpha) \phi U(\alpha)^\dagger,\qquad U(\alpha) =\exp \left( -\frac{\mathrm{i}}{2}\alpha
\sigma _{3}\right) =\mathrm{diag}\left( \mathrm{e}^{-\frac{\mathrm{i}}{2}%
\alpha },\mathrm{e}^{\frac{\mathrm{i}}{2}\alpha }\right) ,
\end{equation}%
which can be rewritten in components as 
\begin{equation}
\phi _{0}\rightarrow \phi _{0},~~~\phi _{\pm }\rightarrow \mathrm{e}^{\pm 
\mathrm{i}\alpha }\phi _{\pm },
\end{equation}
is therefore manifest. Only the charge preserving amplitudes are thus nonzero.

The cubic part of the Lagrangian reads up to integration by parts
\begin{eqnarray}
\mathcal{L}_{\mathrm{cubic}} &=&\lambda _{3}\phi _{0}\left[\frac{1}{6}\left(
\square \phi _{0}\square \phi _{0}-\partial _{\mu }\partial _{\nu }\phi
_{0}\partial ^{\mu }\partial ^{\nu }\phi _{0}\right)  +\left( \square \phi _{+}\square \phi _{-}-\partial
_{\mu }\partial _{\nu }\phi _{+}\partial ^{\mu }\partial ^{\nu }\phi
_{-}\right)\right]
\end{eqnarray}
and we have the following Feynman rules for the cubic vertices
\begin{equation}
V_{3}\left( 1,2,3\right) _{000}=V_{3}\left( 1,2,3\right) _{0+-}=\lambda
_{3}G\left( 1,2\right) =\lambda _{3}G\left( 1,3\right) =\lambda _{3}G\left(
2,3\right)\, .
\end{equation}
Inserting this special values into the general soft theorem we find for the
neutral soft particle
\begin{eqnarray}
&&A_{n+1}\left( tp^{0},1(t) ^{+},\ldots ,k(t)
^{+},\left( k+1\right) (t) ^{-},\ldots ,\left( 2k\right) (t) ^{-},\left( 2k+1\right) (t) ^{0},\ldots n(t) ^{0}\right)    \notag
\\
&&\overset{t\rightarrow 0}{=}O\left( t^{2}\right).
\label{neutral_soft_theorem}
\end{eqnarray}
For the positively charged soft particle we get
\begin{align}
&A_{n+1}\left( tp^{+},1(t)^{+},\ldots , k(t) ^{+},(k+1)(t) ^{-},\ldots ,(2k+1) (t) ^{-},(2k+2)(t)^{0},\ldots n(t) ^{0}\right)   \notag \\
&\overset{t\rightarrow 0}{=}t\frac{\lambda _{3}}{2}\sum_{j=2k+2}^{n}(p\cdot j) A_{n}\left( j^{+},1^{+},\ldots , k^{+},(k+1)^{-},\ldots ,(2k+1) ^{-}, (2k+2) ^{0},\ldots ,\widehat{j^{0}},\ldots ,n^{0}\right)   \notag \\
&+t\frac{\lambda _{3}}{2}\sum_{j=k+1}^{2k+1}(p\cdot j)
A_{n}\left( 1^{+},\ldots , k ^{+},(k+1)^{-},\ldots  ,%
\widehat{j^{-}},\ldots ,\left( 2k+1\right) ^{-},\left( 2k+2\right) ^{0},\ldots
,n^{0},j^{0}\right)   \notag \\
&+O\left( t^{2}\right) ,
\label{U(2)/U(1)_soft_theorem}
\end{align}
and similarly for the negatively charged one. Note that this soft theorem is a 
$\rho =2$ analog of the soft theorem for the $U(2)/U(1)$ nonlinear sigma model \cite{Kampf:2019mcd}\footnote{However, on the contrary to the fibrated $CP^1$ sigma model, the soft theorem (\ref{U(2)/U(1)_soft_theorem}) together with the $U(1)$ symmetry does not select the $U(2)/U(1)$ Galileon uniquely. 
Indeed, there are in fact three-parametric family of 4pt Galileon vertices (and in general $[n/2]+1$ parametric family of $n$-pt Galileon vertices, $n\ge 4$) which are  invariant with respect to the $U(1)$ transformations.
These can be included in the Lagrangian instead of those chosen in $\mathcal{L}_{U(2)/U(1)}$ without changing the particular form (\ref{U(2)/U(1)_soft_theorem}), (\ref{neutral_soft_theorem}) of the soft theorems.}.
To avoid confusion let us also stress that the name $U(2)/U(1)$ we have picked for this model  does not correspond to a `standard' $G/H$-coset construction since the Lagrangian of this theory is not invariant under nonlinearly realized $U(2)$ transformations.

\section{Dualities of multi-flavor Galileon theories}\label{section_duality}

The nontrivial soft theorem of the form (\ref{soft_theorem_galileon}) for multi-flavor Galileon theory is possible only when there are cubic vertices in the Lagrangian. 
However, this is not a sufficient condition for nonvanishing right-hand side of eq. (\ref{soft_theorem_galileon}). 
The $O(t^{2})$ behavior of the full amplitudes is possible also in the theories with cubic vertices, provided these can be removed by a field redefinition without changing the form (up to a change of the couplings $\lambda _{i_{1}\ldots i_{n}}$) of the other vertices of the Lagrangian. We have then a duality between multi-flavor Galileon theory with cubic couplings and some other multi-flavor Galileon theory without cubic couplings.\footnote{For single-Galileon case, the duality of this type has been revealed in \cite{deRham:2013hsa} and further discussed in \cite{Kampf:2014rka,deRham:2014lqa,Creminelli:2014zxa}, while the study of the multi-Galileon case has been initiated in \cite{Noller:2015eda}.}
In this section we will discuss the necessary and sufficient conditions for the existence of such a duality.

\subsection{The necessary conditions -- the amplitude approach}

Let us assume two $N$-flavor Galileon theories with Lagrangians  ${\cal{L}}^{I}$, $I=1$ (with cubic vertices) and $I=2$ (without cubic vertices) which are dual to each other in the sense mentioned above.
Under such a duality, the $S-$matrix is not changed. 
Therefore, the necessary condition for the existence of the duality is the equality of the $n$-point amplitudes in both theories.  
Let us compare first the contribution to the 4pt amplitudes in both cases. 
The contributions of 4pt vertices give rise to the contact terms
\begin{eqnarray}
A_{C}^{\left( I\right) }(1^{a},2^{b},3^{c},4^{d}) &=&\lambda _{abcd}^{\left(
I\right) }G\left( 1,2,3\right) =2\lambda _{abcd}^{\left( I\right) }\left(
1\cdot 2\right) \left( 1\cdot 3\right) \left( 2\cdot 3\right)  \notag \\
&=&\frac{2}{3}\lambda _{abcd}^{\left( I\right) }\left[ \left( 1\cdot
2\right) ^{3}+\left( 1\cdot 3\right) ^{3}+\left( 2\cdot 3\right) ^{3}\right]\,.
\label{contact_4pt}
\end{eqnarray}
In the theory with cubic vertices ($I=1$) we have also contribution of the one-propagator graphs\footnote{Note the summation over repeated index $l$.}
\begin{eqnarray}
A_{P}^{\left( 1\right) }(1^{a},2^{b},3^{c},4^{d}) 
&=&-\lambda _{abl}^{\left( 1\right) }\lambda _{cdl}^{\left( 1\right) }G\left(
1\cdot 2\right)\frac{1}{(1+2)^{2}}G\left( 3\cdot 4\right)+\text{cross} \notag \\
&=&-\frac{1}{2}
\left[ \lambda _{abl}^{\left( 1\right) }\lambda
_{cdl}^{\left( 1\right) }\left( 1\cdot 2\right) ^{3}+\lambda _{acl}^{\left(
1\right) }\lambda _{bdl}^{\left( 1\right) }\left( 1\cdot 3\right)
^{3}+\lambda _{adl}^{\left( 1\right) }\lambda _{bcl}^{\left( 1\right)
}\left( 2\cdot 3\right) ^{3}\right]\,.
\end{eqnarray}
The duality requires  
$
A^{\left( 2\right) }(1^{a},2^{b},3^{c},4^{d})=A^{\left( 1\right)
}(1^{a},2^{b},3^{c},4^{d}),
$
or more explicitly%
\begin{eqnarray}
&&\frac{2}{3}\left[\lambda _{abcd}^{\left( 2\right) }-\lambda _{abcd}^{\left( 1\right) }\right]\left[ \left( 1\cdot 2\right)
^{3}+\left( 1\cdot 3\right) ^{3}+\left( 2\cdot 3\right) ^{3}\right]\notag\\
&&\;\;\;\;\;\;\;\;\;\;\;\;\;=-\frac{1}{2}\left[ \lambda _{abl}^{\left( 1\right) }\lambda _{cdl}^{\left(
1\right) }\left( 1\cdot 2\right) ^{3}+\lambda _{acl}^{\left( 1\right)
}\lambda _{bdl}^{\left( 1\right) }\left( 1\cdot 3\right) ^{3}+\lambda
_{adl}^{\left( 1\right) }\lambda _{bcl}^{\left( 1\right) }\left( 2\cdot
3\right) ^{3}\right]\,,
\end{eqnarray}%
which is possible only\footnote{%
This can be easily proved by means of expressing both contributions in terms
of two independent variables (say $(1\cdot 2)$ and $(1\cdot 3)$) and
comparing the coefficients of the corresponding polynomials.} when for any
permutation $\sigma \in S_4$%
\begin{equation}
\lambda _{\sigma (a)\sigma (b)l}^{\left( 1\right) }\lambda _{\sigma
(c)\sigma (d)l}^{\left( 1\right) }=\frac{4}{3}\left[ \lambda _{abcd}^{\left(
1\right) }-\lambda _{abcd}^{\left( 2\right) }\right]\,.
\label{permutational_symmetry}
\end{equation}%
As a consequence of the total symmetry of $\lambda _{abcd}^{(I)}$, the
sum $\lambda_{abl}^{(1)}\lambda_{cdl}^{(1)}$ must be also totally symmetric.

Similarly, the 5pt amplitudes in both theories should coincide, i.e. 
$A^{(1)}(1^{a},2^{b},3^{c},4^{d},5^{e})$ $= A^{(2)}(1^{a},2^{b},3^{c},4^{d},5^{e})$.
In the theory $I=2$ we get just the contribution of the contact term
\begin{equation}
A_{C}^{\left( I\right) }(1^{a},2^{b},3^{c},4^{d},5^{e})=\lambda
_{abcde}^{\left( I\right) }G\left( 1,2,3,4\right)\,,
\end{equation}
while in the case $I=1$ we have also the one propagator terms
\begin{eqnarray}
A_{1P}^{\left( 1\right) }(1^{a},2^{b},3^{c},4^{d},5^{e}) &=&-\lambda
_{abl}^{\left( 1\right) }\lambda _{cdel}^{\left( 1\right) }G\left(
1,2\right) \frac{1}{(1+2)^{2}}G\left( 3,4,5\right) +\text{cross} \notag\\
&=&\lambda _{abl}^{\left( 1\right) }\lambda _{cdel}^{\left( 1\right) }\left(
1\cdot 2\right) \left( 3\cdot 4\right) \left( 4\cdot 5\right) (3\cdot
5)+\text{cross}
\end{eqnarray}
and the two propagator terms\footnote{Here we sum over $k,l$.} 
\begin{eqnarray}
A_{2P}^{\left( 1\right) }(1^{a},2^{b},3^{c},4^{d},5^{e}) &=&\lambda
_{abk}^{\left( 1\right) }\lambda _{ckl}^{\left( 1\right) }\lambda
_{del}^{\left( 1\right) }  G\left( 1,2\right) \frac{1}{(1+2)^{2}}G(1+2,4+5)\frac{1}{(4+5)^{2}}%
G\left( 4,5\right) \notag\\
&&+\text{cross} \notag\\
&=&-\frac{1}{4}\lambda _{abk}^{\left( 1\right) }\lambda _{ckl}^{\left(
1\right) }\lambda _{del}^{\left( 1\right) }\left( 1\cdot 2\right) \left(
4\cdot 5\right) \left[ \left( 1+2\right) \cdot 3\right] ^{2} +\text{cross}\,.
\end{eqnarray}
The necessary condition for the existence of the duality reads then 
\begin{eqnarray}
\left( \lambda _{abcde}^{\left( 2\right) }-\lambda _{abcde}^{\left(
1\right) }\right) G\left( 1,2,3,4\right)
&=&\lambda _{abl}^{\left( 1\right) }\lambda _{cdel}^{\left( 1\right) }\left(
1\cdot 2\right) \left( 3\cdot 4\right) \left( 4\cdot 5\right) (3\cdot 5)\notag\\&-&%
\frac{1}{4}\lambda _{abk}^{\left( 1\right) }\lambda _{ckl}^{\left( 1\right)
}\lambda _{del}^{\left( 1\right) }\left( 1\cdot 2\right) \left( 4\cdot
5\right) \left[ \left( 1+2\right) \cdot 3\right] ^{2}+\text{cross}\,.
\label{comparison}
\end{eqnarray}
Note that
\begin{equation}
\left( 1\cdot 2\right) \left( 3\cdot 4\right) \left( 4\cdot 5\right) (3\cdot
5)+\text{cross}=-2G\left( 1,2,3,4\right)
\end{equation}
and the above condition is fulfilled only\footnote{Again, this can be proved by means of expressing the amplitudes in terms of the five independent scalar products and comparing the coefficients of the polynomials on both sides of (\ref{comparison}).} when for
any permutation $\sigma\in S_5 $
\begin{eqnarray}
\lambda _{\sigma (a)\sigma (b)l}^{\left( 1\right) }\lambda _{\sigma
(c)\sigma (d)\sigma (e)l}^{\left( 1\right) } &=&\lambda _{abl}^{\left(
1\right) }\lambda _{cdel}^{\left( 1\right) }\notag \\
\lambda _{\sigma \left( a\right) \sigma \left( b\right) k}^{\left( 1\right)
}\lambda _{\sigma \left( c\right) kl}^{\left( 1\right) }\lambda _{\sigma
\left( d\right) \sigma \left( e\right) l}^{\left( 1\right) } &=&2\lambda
_{abl}^{\left( 1\right) }\lambda _{cdel}^{\left( 1\right) }+\left( \lambda
_{abcde}^{\left( 2\right) }-\lambda _{abcde}^{\left( 1\right) }\right) .
\end{eqnarray}
The tensor $\lambda _{abl}^{\left( 1\right) }\lambda _{cdel}^{\left(
1\right) }$ must be therefore totally symmetric. 
Note that the double sum 
$
\lambda _{abk}^{\left( 1\right) }\lambda _{ckl}^{\left( 1\right) }\lambda
_{del}^{\left( 1\right) }
$ 
is totally symmetric automatically as a consequence of (\ref{permutational_symmetry}). The second condition above, which fixes the difference $\lambda
_{abcde}^{\left( 2\right) }-\lambda _{abcde}^{\left( 1\right) }$, is thus
consistent.

Similar constraints can be derived  for higher $n$-point amplitudes for $n\le D+1$. 
These constraints give a set of \emph{necessary} conditions for the existence of the duality. 
Analogously as above we can formulate these conditions as a requirement that all for $m\leq D$ the tensors in the flavor space  $\lambda_{i_{1}\ldots i_{m-1}k}\lambda_{i_{m}i_{m+1}k}$ are totally symmetric.

\subsection{The sufficient conditions -- the Lagrangian approach}

Let us now discuss the sufficient conditions for the existence of the
duality. 
We can write the cubic vertex as (cf.~(\ref{Lagrangian_basic_alternative}))
\begin{equation*}
\mathcal{L}_{N}^{\left( 3\right) }=-\frac{1}{4}\lambda _{ijk}\left( \partial
\phi _{i}\cdot \partial \phi _{j}\right) \square \phi _{k}
\end{equation*}%
and assume the following infinitesimal field reparameterization%
\begin{equation}
\delta _{\theta }\phi _{k}=-\frac{1}{4}\theta \lambda _{ijk}\left( \partial
\phi _{i}\cdot \partial \phi _{j}\right) .  \label{duality_transformation}
\end{equation}%
Then the change of the kinetic term reads (up to the integration by parts)%
\begin{equation*}
\delta _{\theta }\int \mathrm{d}^{D}x\frac{1}{2}\partial \phi _{k}\cdot
\partial \phi _{k}=\int \mathrm{d}^{D}x\frac{1}{4}\theta \lambda
_{ijk}\left( \partial \phi _{i}\cdot \partial \phi _{j}\right) \square \phi
_{k}.
\end{equation*}
This effectively shifts
\begin{equation*}
\lambda _{ijk}\rightarrow \left( 1-\theta \right) \lambda _{ijk}
\end{equation*}
and thus iteration of this transformation can be used to eliminate the cubic vertex. 
However, the duality mentioned above means more than this. 
Namely, we need that the form of the action is not changed (up to a shift of the couplings). 
The change of the general terms of the action is\footnote{Note that $\delta_{\theta}\int {\rm d}^{D}x{\cal{L}}^{(D+1)}_{N}=0$ in $D$ dimensions.}
\begin{eqnarray}\label{Ln_transformation}
\delta _{\theta }\int \mathrm{d}^{D}x\mathcal{L}_{N}^{\left( n\right) } &=&%
\frac{\left( -1\right) ^{n-1}}{n!}\lambda _{i_{1}\ldots i_{n}}\delta
_{\theta }\int \mathrm{d}^{D}x\phi _{i_{n}}\delta _{\nu _{1}\ldots \nu
_{n-1}}^{\mu _{1}\ldots \mu _{n-1}}\prod\limits_{l=1}^{n-1}\partial _{\mu
_{l}}\partial^{\nu _{l}}\phi _{i_{l}}   \\
&=&\frac{\left( -1\right) ^{n}}{4\left( n-1\right) !}\theta \lambda
_{i_{1}\ldots i_{n-1}k}\lambda _{ijk}\int \mathrm{d}^{D}x\left( \partial
\phi _{i}\cdot \partial \phi _{j}\right) \delta _{\nu _{1}\ldots \nu
_{n-1}}^{\mu _{1}\ldots \mu _{n-1}}\prod\limits_{l=1}^{n-1}\partial _{\mu
_{l}}\partial ^{\nu _{l}}\phi _{i_{l}}\,.  \notag
\end{eqnarray}
However, as follows from  (\ref{Lagrangian_basic}) and (\ref{Lagrangian_basic_alternative}), provided $\lambda _{i_{1}\ldots i_{n-1}k}\lambda
_{i_{n}i_{n+1}k}$ is totally symmetric, i.e. when for any permutation $\sigma\in S_{n+1}$ and for $n\le D$
\begin{equation}
\lambda _{i_{1}\ldots i_{n-1}k}\lambda _{i_{n}i_{n+1}k}=\lambda _{i_{\sigma
\left( 1\right) }\ldots i_{\sigma \left( n-1\right) }k}\lambda _{i_{\sigma
(n)}i_{\sigma (n+1)}k},  \label{condition}
\end{equation}
the transformation formula (\ref{Ln_transformation}) effectively means a
shift\footnote{Cf. the discussion of the equivalent form of the multi-flavor Galileon Lagrangians after eq.~(\ref{Lagrangian_single_alternative}).} of the coupling $\lambda _{i_{1}\ldots i_{n+1}}$
\begin{equation*}
\lambda _{i_{1}\ldots i_{n+1}}\rightarrow \lambda _{i_{1}\ldots
i_{n+1}}-n\theta \lambda _{i_{1}\ldots i_{n-1}k}\lambda _{i_{n}i_{n+1}k}.
\end{equation*}
Thus, provided (\ref{condition}) is fulfilled, the iterations of the infinitesimal transformation (\ref{duality_transformation}) is a desired duality. 
Note that these sufficient conditions are the same as the necessary conditions derived in the previous subsection.

\subsection{Solution of the necessary and sufficient conditions\label{solution_of_necessary_condition}}

As we have found in the previous subsections, the necessary and sufficient conditions for the existence of a duality transformation which allows to remove the cubic couplings are given by (\ref{condition}).
These conditions are (due to the symmetry of $\lambda _{ijk}$ and $\lambda_{i_{1}\ldots i_{n}}$) equivalent to
\begin{equation}
\lambda _{i_{1}\ldots i_{n-1}k}\lambda _{i_{n}i_{n+1}k}=\lambda
_{i_{1}\ldots i_{n}k}\lambda _{i_{n-1}i_{n+1}k}.
\label{modified_condition}
\end{equation}
for $n\le D$.
In this subsection we give a general solution of these relations (up to an $SO(N)$ rotation in the flavor space).

Let us start with the case $n=3$, for which we get
\begin{equation}
\lambda _{ail}\lambda _{blj}=\lambda _{bil}\lambda _{alj}\,.
\end{equation}
Defining the $N\times N$ symmetric matrices $\Lambda^{(a)}$, $a=1,\ldots ,N$ with matrix elements $\Lambda _{ij}^{(a)}=\lambda_{aij}$, this can be rewritten as
\begin{equation}
\bigl[ \Lambda ^{\left( a\right) },\Lambda ^{\left( b\right) }\bigr] =0\,.
\end{equation}
This means, that $\Lambda^{(a)}$s can be diagonalized simultaneously by $SO(N)$ matrix $M$
\begin{equation}
M_{im}M_{jn}\Lambda _{ij}^{(a) }=\alpha _{m}^{(a)}\delta _{mn}
\end{equation}
(there is no summation over $m$). Let us define also
\begin{equation}
\beta _{m}^{(a)}=M_{ba}\alpha _{m}^{(b)}.
\end{equation}
After field redefinition
\begin{equation}
\phi _{i}=M_{ij}\psi _{j}\,,
\end{equation}
we get for the cubic vertex in terms of the new fields
\begin{eqnarray}
\mathcal{L}_{N}^{\left( 3\right) } &=&-\frac{1}{4}\lambda _{kij}\left(
\partial \phi _{i}\cdot \partial \phi _{j}\right) \square \phi _{k} 
=-\frac{1}{4}\Lambda _{ij}^{\left( k\right) }M_{im}M_{jn}M_{kl}\left(
\partial \psi _{m}\cdot \partial \psi _{n}\right) \square \psi _{l} \notag\\
&=&-\frac{1}{4}\sum_{l,m,n=1}^{N}\beta _{m}^{\left( l\right) }\delta
_{mn}\left( \partial \psi _{m}\cdot \partial \psi _{n}\right) \square \psi
_{l}\,.
\end{eqnarray}
Since 
$
\beta _{m}^{\left( l\right) }\delta _{mn}=M_{im}M_{jn}M_{kl}\lambda _{kij}
$ 
and $\lambda _{kij}$ is totally symmetric, we get 
 $
\beta _{m}^{\left( l\right) }\delta _{mn}=\beta _{m}^{\left( n\right)
}\delta _{ml}
$ 
and for $m=n$
\begin{equation}
\beta _{m}^{\left( l\right) }=\beta _{m}^{\left( m\right) }\delta
_{ml}\equiv \gamma _{m}\delta _{ml}\,.
\end{equation}
Finally we have (here no summation over $m$ is understood)
\begin{equation}
\beta _{m}^{\left( l\right) }\delta _{mn}=\gamma _{m}\delta _{ml}\delta _{mn}
\end{equation}
and as a result
\begin{equation}
\mathcal{L}_{N}^{\left( 3\right) } =-\frac{1}{4}\sum_{l,m,n=1}^{N}\gamma
_{m}\delta _{ml}\delta _{mn}\left( \partial \psi _{m}\cdot \partial \psi
_{n}\right) \square \psi _{l}
=-\frac{1}{4}\sum_{m=1}^{N}\gamma _{m}\left( \partial \psi _{m}\cdot
\partial \psi _{m}\right) \square \psi _{m}.
\end{equation}
Therefore, in terms of the new fields the cubic interaction is decoupled in the
flavor space.

Let $\lambda _{i_{1}\ldots i_{n}}$ ($n>3$) be the corresponding
higher point couplings in the new field basis.
The necessary and sufficient conditions (\ref{modified_condition}) for the
existence of the duality are then 
\begin{equation}
\lambda _{i_{1}\ldots i_{n-2}jk}\gamma _{m}\delta _{ml}\delta _{mk}=\lambda
_{i_{1}\ldots i_{n-2}lk}\gamma _{m}\delta _{mj}\delta _{mk}
\end{equation}%
or
\begin{equation}
\lambda _{i_{1}\ldots i_{n-2}jm}\gamma _{m}\delta _{ml}=\lambda
_{i_{1}\ldots i_{n-2}lm}\gamma _{m}\delta _{mj}\,.  \label{solution}
\end{equation}%
For $j=m$ this means (no summation over $j$) 
\begin{equation}
\lambda _{i_{1}\ldots i_{n-2}jj}\gamma _{j}\delta _{jl}=\lambda
_{i_{1}\ldots i_{n-2}lj}\gamma _{j}\,,
\end{equation}%
i.e. either $\gamma _{j}=0$ or%
\begin{equation}
\lambda _{i_{1}\ldots i_{n-2}lj}=\lambda _{i_{1}\ldots i_{n-2}jj}\delta _{jl}\,.
\end{equation}%
Total symmetry of $\lambda _{i_{1}\ldots i_{n}}$ gives then in this case for 
$k=1,\ldots ,n-2$,%
\begin{equation}
\lambda _{i_{1}\ldots i_{n-2}lj}=\lambda _{i_{1}\ldots l\ldots
i_{n-2}jj}\delta _{ji_{k}}
\end{equation}%
and thus $\lambda _{i_{1}\ldots i_{n-2}lj}=0$ unless $i_{1}=i_{2}=\ldots
=i_{n-2}=j$. Therefore whenever $\gamma _{j}\neq 0$ 
\begin{equation}
\lambda _{i_{1}\ldots i_{n-2}i_{n-1}j}=\rho ^{\left( j\right) }\delta
_{i_{1}j}\delta _{i_{2}j}\ldots \delta _{i_{n-2}j}\delta _{i_{n-1}j}\,.
\end{equation}%
Therefore, we get the following general picture: Let us divide the set $%
\left\{ 1,2,\ldots ,N\right\} $ into two disjoint subsets corresponding to
different values of $\gamma _{j}$%
\begin{equation}
\left\{ 1,2,\ldots ,N\right\} =I_{0}\cup I_{1}\,,
\end{equation}%
where $j\in I_{0}$ iff $\gamma _{j}=0$ and $j\in I_{1}$ when $\gamma
_{j}\neq 0$. Then for 
\begin{equation}
i_{1},i_{2},\ldots ,i_{n}\in I_{0}
\end{equation}%
the $\lambda _{i_{1}i_{2}\ldots i_{n}}$ are arbitrary (totally symmetric).
For 
\begin{equation}
i_{1},i_{2},\ldots ,i_{n}\in I_{1}
\end{equation}%
we get%
\begin{equation}
\lambda _{i_{1}\ldots i_{n}}=\rho ^{\left( i_{n}\right) }\delta
_{i_{1}i_{n}}\delta _{i_{2}i_{n}}\ldots \delta _{i_{n-1}i_{n}}
\end{equation}%
and 
\begin{equation}
\lambda _{i_{1}\ldots i_{n}}=0\,,
\end{equation}%
if at least one $i_{k}\in I_{0}$ and at least one $i_{r}\in I_{1}$.

To summarize, the 3pt vertices can be removed by Galileon duality transformation when and only when the Lagrangian (\ref{Lagrangian_basic}) of the theory becomes after an appropriate $SO(N)$ rotation in the flavor space  a sum of a multi-Galileon theory with $\left\vert I_{0}\right\vert $ flavors without cubic vertices and $N-\left\vert I_{0}\right\vert$ single Galileon theories without any restriction on the presence of cubic vertices.
Note that such a theory obeys the enhanced Adler zero with soft exponent $\sigma=2$.

\section{Reconstruction of the multi-Galileon amplitudes via soft theorem}\label{sec:5}

Up to now we have defined the multi-flavor Galileon theories by means of introducing the Lagrangian (\ref{Lagrangian_basic}) and identifying the corresponding symmetry, which implied the validity of the soft theorem of the general form (\ref{soft_theorem_galileon}).
Though we obtained some information on the conditions under  which the multi-flavor Galileons obey the enhanced Adler zero with soft exponent $\sigma=2$ in the previous section, we have still not proved  that the theories satisfying these conditions are the only ones which have this property. 
In this section we will therefore discuss an alternative definition for the multi-flavor Galileon theories based solely on the soft theorem and the Galileon power counting.
An appropriate tool for such a discussion is a soft BCFW recursion \cite{Cheung:2015ota}, which we will generalize to the multi-flavor case.

For simplicity, in  what follows we will restrict ourselves to $D=4$ dimensions, unless stated otherwise.

\subsection{Soft BCFW recursion for multi-flavor Galileon}

It is known that the tree-level amplitudes of the single Galileon theories are uniquely reconstructible from the seed amplitudes.
This means that in four dimensions, once all the tree-level amplitudes up to five-point\footnote{In general $D$ dimensions we have to fix all the amplitudes up to  $(D+1)$-point one.} are fixed, the remaining tree amplitudes can be uniquely obtained by means of an appropriate recursive procedure.
The most systematic one for such a purpose is the soft BCFW recursion developed in \cite{Cheung:2015ota}. 
It takes into account the one-particle unitarity as well as the defining property of the theory, namely the power counting  $\rho=2$ and the enhanced soft limit with the soft exponent $\sigma=2$.
This procedure can be easily generalized for  the multi-flavor case and for the general soft theorem of the form (\ref{soft_theorem_galileon}).

Indeed, for an $n$-point amplitude $A_{n}( p_{1}^{i_1},\ldots ,p_{n}^{i_n}) $ let us introduce the usual 
deformation of the momenta, which allows to probe the single soft limits of all the external particles
\begin{equation}
p_{i}\rightarrow p_{i}(z) =\left( 1-a_{i}z\right) p_{i}.
\label{deformation}
\end{equation}
Here the parameters $a_{i}$ are constrained according to the momentum conservation 
\begin{equation}
\sum_{i=1}^{n}a_{i}p_{i}=0.
\label{deformation_constraint}
\end{equation}
For $n>D+1$ there exists a nonzero $(n-D)$-parametric solution for $a_k$. For instance in the special case $D=4$, $n=6$ we can write the solution in the form (under the assumption that
arbitrary selection of four out of the six momenta is linearly
independent)
\begin{equation}
a_{k}=a_{i}+\left( -1\right) ^{j-k}\mathrm{sign}\left( k-i\right) \mathrm{%
sign}\left( j-i\right) \left( a_{j}-a_{i}\right) \frac{\left. \det \left(
p_{1},\ldots ,p_{6}\right) \right\vert _{i,k}}{\left. \det \left(
p_{1},\ldots ,p_{6}\right) \right\vert _{i,j}}
\end{equation}
for $k\neq i,j$ while $a_{i}$ and $a_{j}$ are arbitrary. 
Here $\det(\cdot)\vert _{ij}$ is a determinant of the matrix $%
\left( p_{1},\ldots ,p_{6}\right) $ with $i$-th and $j$-th column excluded,
e.g.
\begin{equation}
\left. \det \left( p_{1},\ldots ,p_{6}\right) \right\vert _{2,5}=\det \left(
p_{1},p_{3},p_{4},p_{6}\right)\,.
\end{equation}
The solution is therefore two-parametric, these parameters correspond to the
simultaneous shift (by $a_{i}$) of all $a_{k}$'s and simultaneous re-scaling
(by $\left( a_{j}-a_{i}\right) $) of all $a_{k}$'s where $k\ne i,j$. 
The solution can be
rewritten in more symmetric form as
\begin{equation}
a_{k}-a_{i}=\alpha \left( -1\right) ^{k-i}\mathrm{sign}\left( k-i\right)
\left. \det \left( p_{1},\ldots ,p_{6}\right) \right\vert _{i,k}
\label{a_difference}
\end{equation}
valid for all $k\neq i=1,\ldots 6$ where $\alpha $ is a free parameter. 

The reconstruction of the $n$-point amplitude from the lower-point amplitudes corresponds to the application of the residue theorem to the function $f(z)/z$ where
\begin{equation}
f(z) =\frac{A_{n}\bigl( p_{1}^{i_1}( z) ,\ldots
,p_{n}^{i_n}( z) \bigr) }{\prod\limits_{j=1}^{n}( 1-a_{j}z)
^{2}}\overset{z\rightarrow \infty }{=}O( z^{-2}).
\label{f_asymptotics}
\end{equation}
The indicated asymptotic behavior is a consequence of the power counting $\rho=2$.
The function $f(z)$ has both the unitarity poles $z=z^{\pm}_{{\cal{F}}}$ and the additional poles for $z=1/a_{j}$. 
We get therefore
\begin{equation}
A_{n}\bigl( p_{1}^{i_1},\ldots ,p_{n}^{i_n}\bigr) =\mathrm{res}\Bigl( \frac{f(z)}{z},0\Bigr) =-\sum_{\mathcal{F},I=\pm}\mathrm{res}\Bigl( \frac{f(z)}{z},z_{\mathcal{F}}^{I}\Bigr)
-\sum_{j=1}^{n}\mathrm{res}\Bigl( \frac{f(z)}{z},\frac{1}{a_j}\Bigr)\,.
\label{BCFW}
\end{equation}
Here $z_{\mathcal{F}}^{\pm }$ are two solutions of the quadratic equation
corresponding to the factorization channel $\mathcal{F}$
\begin{equation}
p_{\mathcal{F}}^{2}(z) \equiv \biggl(\sum_{s\in \mathcal{F}}p_{s}(z)\biggr)^2=0\,.
\end{equation}
The residue at $z=z_{\mathcal{F}}^{\pm }$ is determined from unitarity and
reads%
\begin{equation}
-\mathrm{res}\left( \frac{f(z) }{z},z_{\mathcal{F}}^{\pm
}\right) =\sum_{i_{\cal{F}}}\frac{A_{L}\left( \mathcal{F}\left( z_{\mathcal{F}}^{\pm };\right)i_{\cal{F}}
\right) A_{R}\left( \overline{\mathcal{F}}\left( z_{\mathcal{F}}^{\pm
}\right) ;\overline{i_{\cal{F}}}\right) }{z_{\mathcal{F}}^{\pm }\prod\limits_{j=1}^{n}\left(
1-a_{j}z_{\mathcal{F}}^{\pm }\right) ^{2}}\mathrm{res}\left( \frac{1}{p_{%
\mathcal{F}}^{2}(z) },z_{\mathcal{F}}^{\pm }\right),
\label{unitarity_pole}
\end{equation}
where the sum on the right-hand side is over all  particle flavors $i_{\cal{F}}$ which can be exchanged in the factorization channel ${\cal{F}}$.
The remaining residues at $z=1/a_i$  either vanish when the theory has enhanced Adler zero with soft exponent $\sigma=2$, or can be straightforwardly determined from the generalized soft theorem (\ref{generalized_soft_theorem}) as
\begin{equation}
 -\mathrm{res}\left( \frac{f(z) }{z},\frac{1}{a_i}\right) =  \sum_{k\ne i}\frac{\lambda_{i_{i}i_{k}l}\left(p_i\cdot p_k(1/a_i)\right)A_{n-1}^{i}(p_{1}^{i_1}(1/a_i),\dots,p_{k}^{l}(1/a_i),\dots,p_{n}^{i_n}(1/a_i))}{2\prod\limits_{j\ne i}\bigl(
1-\frac{a_{j}}{a_{i}}\bigr)^2}\,, 
\label{third_term}
\end{equation}
where $A_{n-1}^{i}$ is the $(n-1)$-point amplitude with the $i$-th particle omitted from the original $n$-point configuration.

All the residues are therefore expressed in terms of the lower-point amplitudes and the residue theorem (\ref{BCFW}) thus gives the desired modified soft BCFW recursion relation.
Since the  momentum deformation (\ref{deformation}) is possible only for $n\ge 6$ in four dimensions (or with $n\ge D+2$ in general $D$ dimensions), we need the above mentioned set of seed amplitudes $A_k$, $k=4,\dots (D+1)$ as the initial conditions for the recursion. 

Provided we have fixed the general multi-flavor Galileon Lagrangian (\ref{Lagrangian_basic}), the seed amplitudes can be calculated uniquely in terms of the couplings $\lambda_{i_1\dots i_k}$ and since the theory exists, the recursion can be used as an alternative way for the calculation of the tree-level amplitudes.
On the other hand, if the Lagrangian is not known, we can try to construct the most general set of the seed amplitudes compatible with the prescribed soft theorem and power counting and try to define the theory via the soft BCFW recursion.
However, in such a case we have \emph{a priori} no guaranty that such a theory exists and that the whole procedure makes sense. 
The best we can do is to test the consistency of the first several iterations of the recursion and provided an inconsistency is found, to make a negative statement about the existence of the theory.
As such a consistency check we can use the independence of the BCFW reconstructed $n>D+1$ amplitudes on the free parameters which parametrize the solution $\{a_i\}_{i=1}^{n}$ of the constraints (\ref{deformation_constraint}).  
This strategy is known as the soft bootstrap and has been originally used for the exploration of the landscape of single-flavor effective field theories \cite{Cheung:2016drk} and their possible SUSY extensions  \cite{Elvang:2018dco}.
In the case of the multi-flavor theories with the Galileon power counting we can further simplify the method and instead of exploring the full BCFW construction of the higher-point amplitudes we can use the so-called bonus relations for constraining the seed amplitudes.
The method of bonus relations for six-point amplitudes in $D=4$ will be described in the next subsections.

\subsection{Reconstruction of the 6-point amplitudes}
\label{sec:rec6}
Let us first discuss the general parametrization of the  four-point seed amplitudes.
Since the three-point amplitudes in the theory with Galileon power counting vanish on shell, the most general 4pt amplitude in such a theory is contact and therefore it can be expressed as a polynomial of the third order in two independent Mandelstam variables $s_{12}$ and $s_{13}$, where
\begin{equation}
s_{ij}=\left( p_{i}+p_{j}\right) ^{2}, \,\,\,\,i,j=1,\dots,4.
\end{equation}
The above polynomial is a linear combination of basic monomials 
$
\left\{ s_{12}^{3},~s_{13}^{3},~s_{12}^{2}s_{13},~s_{12}s_{13}^{2}\right\}
$. 
Note that the following identity applies
\begin{equation}
s_{12}^{3}+s_{13}^{3}+s_{23}^{3}=3s_{12}s_{13}s_{23}=-3\left(
s_{12}^{2}s_{13}+s_{12}s_{13}^{2}\right).
\end{equation}
So instead we can take the following set as a basis 
\begin{equation}
\left\{ s_{12}^{3},~s_{13}^{3},~s_{23}^{3},~(s_{12}^{2}s_{13}-s_{12}s_{13}^{2}) +\tfrac{2}{3}(s_{12}^{3}-s_{13}^{3}) \right\}.
\end{equation}
This one is even more symmetric as a consequence of the  relation
\begin{equation}
\left( s_{12}^{2}s_{13}-s_{12}s_{13}^{2}\right) +\frac{2}{3}\left(
s_{12}^{3}-s_{13}^{3}\right) =\frac{1}{3}\left( s_{12}-s_{13}\right) \left(
s_{12}-s_{23}\right) \left( s_{13}-s_{23}\right),
\end{equation}
the right-hand side of which is manifestly totally antisymmetric under permutations of $\left\{ 1,2,3,4\right\} $.
The most general contact 4pt amplitude with Galileon power-counting is
therefore
\begin{eqnarray}
A_{4}\left( 1^{a_{1}},2^{a_{2}},3^{a_{3}},4^{a_{4}}\right) &=&\lambda
_{a_{1}a_{2},a_{3}a_{4}}^{\left( 12\right) }s_{12}^{3}+\lambda
_{a_{1}a_{3},a_{2}a_{4}}^{\left( 13\right) }s_{13}^{3}+\lambda
_{a_{2}a_{3},a_{1}a_{4}}^{\left( 23\right) }s_{23}^{3}\notag \\
&&+\lambda _{a_{1}a_{2}a_{3}a_{4}}^{-}\left( s_{12}-s_{13}\right) \left(
s_{12}-s_{23}\right) \left( s_{13}-s_{23}\right)\,.
\label{eq:A4pt}
\end{eqnarray}
Bose symmetry requires
\begin{eqnarray}
\lambda _{a_{1}a_{2},a_{3}a_{4}}^{\left( 12\right) } =\lambda
_{a_{1}a_{2},a_{3}a_{4}}^{\left( 13\right) }=\lambda
_{a_{1}a_{2},a_{3}a_{4}}^{\left( 23\right) }&\equiv& \lambda
_{a_{1}a_{2},a_{3}a_{4}}^{+}=\lambda _{a_{2}a_{1},a_{3}a_{4}}^{+}=\lambda
_{a_{3}a_{4},a_{1}a_{2}}^{+} \label{lambda_symmetries}\\
\lambda _{a_{\sigma (1)}a_{\sigma (2)}a_{\sigma (3)}a_{\sigma (4)}}^{-} &=&%
\mathrm{sign}(\sigma) \lambda _{a_{1}a_{2}a_{3}a_{4}}^{-}
\end{eqnarray}
and thus we are left with only two independent tensors $\lambda_{a_{1}a_{2}a_{3}a_{4}}^{\pm}$ in the flavor space which parametrize the amplitude
\begin{eqnarray}
A_{4}\left( 1^{a_{1}},2^{a_{2}},3^{a_{3}},4^{a_{4}}\right) &=&\lambda
_{a_{1}a_{2},a_{3}a_{4}}^{+}s_{12}^{3}+\lambda
_{a_{1}a_{3},a_{2}a_{4}}^{+}s_{13}^{3}+\lambda
_{a_{2}a_{3},a_{1}a_{4}}^{+}s_{23}^{3} \notag\\
&&+\lambda _{a_{1}a_{2}a_{3}a_{4}}^{-}\left( s_{12}-s_{13}\right) \left(
s_{12}-s_{23}\right) \left( s_{13}-s_{23}\right).
\label{general_4pt}
\end{eqnarray}
This is the most general seed amplitude which can serve as an input for the first iteration of the soft BCFW recursion and which is necessary for the reconstruction of the six-point amplitude.
Note that as a consequence of the four-point kinematics, the amplitude has automatically even more enhanced Adler zero than required, namely, it behaves as $O(p^3)$ in the single-particle soft limit $p\to 0$.
In the case when this amplitude comes from the Lagrangian (\ref{Lagrangian_basic}), we have $\lambda _{a_{1}a_{2}a_{3}a_{4}}^{-}=0$ and provided there are  no cubic terms, the tensor $\lambda_{a_{1}a_{2},a_{3}a_{4}}^{+}$ is totally symmetric.

Let us now use the general formula (\ref{BCFW}) for a six-point amplitude.
Note that this is possible for $D\le 4$, 
since only then we get a non-trivial solution for the $a_j$'s defining the deformation of the momenta (\ref{deformation}).

In the case of the six-point amplitude in $D=4$ we can express the residues at the unitarity poles $z_{\mathcal{F}}^{\pm}$ as
\begin{equation}
-\mathrm{res}\left( \frac{f(z) }{z},z_{\mathcal{F}}^{\pm
}\right) =\mathrm{res}\left( \frac{f_{\mathcal{F}}(z) }{z},z_{%
\mathcal{F}}^{\pm }\right).
\label{modified_residue}
\end{equation}
Here we have defined a new function, schematically (without explicit flavor indices, cf. (\ref{unitarity_pole}))
\begin{equation}
f_{\mathcal{F}}(z) =\sum_{i_{\cal{F}}}\frac{\widetilde{A}_{4,L}\left( \mathcal{F}%
(z);i_{\cal{F}} \right) \widetilde{A}_{4,R}\left( \overline{\mathcal{F}}(z);\overline{i_{\cal{F}}} \right) }{p_{\mathcal{F}}^{2}(z)
\prod\limits_{j=1}^{6}\left( 1-a_{j}z\right) ^{2}}
\label{modified_f}
\end{equation}
and $\widetilde{A}_{4,L}\left( \mathcal{F}(z); i_{\cal{F}} \right)$ and $\widetilde{A}_{4,R}\left( \overline{\mathcal{F}}(z); \overline{i_{\cal{F}}}\right)$ are appropriately chosen analytic continuations of ${A}_{4,L}\left( \mathcal{F}(z^{\pm}_{\cal{F}}); i_{\cal{F}} \right)$ and ${A}_{4,R}\left( \overline{\mathcal{F}}(z^{\pm}_{\cal{F}}); \overline{i_{\cal{F}}} \right)$ (see the explicit formula (\ref{modified_4pt}) below).
These are assumed to respect the Galileon power counting, so that the new function $f_{\mathcal{F}}(z)$ has the same $z\to \infty$ asymptotics as $f(z)$. Using the residue theorem once more, now for the function $f_{\mathcal{F}}(z)$, we can write
\begin{equation}
-\mathrm{res}\left( \frac{f(z) }{z},z_{\mathcal{F}}^{\pm
}\right) =\mathrm{res}\left( \frac{f_{\mathcal{F}}(z) }{z},z_{%
\mathcal{F}}^{\pm }\right) =-\sum_{j=1}^{6}\mathrm{res}\left( \frac{f_{%
\mathcal{F}}(z) }{z},\frac{1}{a_{j}}\right) -f_{\mathcal{F}%
}(0)
\end{equation}
and thus according to (\ref{BCFW})
\begin{equation}
A_{6}\left( p_{1}^{i_1},\ldots ,p_{n}^{i_n}\right) =-\sum_{\mathcal{F}}f_{\mathcal{F}%
}(0) -\sum_{\mathcal{F}}\sum_{j=1}^{6}\mathrm{res}\left( \frac{%
f_{\mathcal{F}}(z) }{z},\frac{1}{a_{j}}\right)-\sum_{j=1}^{6}\mathrm{res}\left( \frac{
f(z)}{z},\frac{1}{a_j}\right).
\label{modified_6pt_recursion}
\end{equation}
We need therefore to calculate only the residues at $z=1/a_j$ which is much easier than the calculation of the residues at the unitarity poles.

Let us now find the explicit form of the functions $f_{\mathcal{F}}(z)$.
There are in principle infinitely many possibilities how to construct the modified four-point amplitudes  $\widetilde{A}_{4}\left( \mathcal{F}(z);i_{\cal{F}} \right)$ which satisfy (\ref{modified_residue}) and (\ref{modified_f}), here we will chose the one which makes the calculation of the residues at $z=1/a_j$ straightforward.

Note that since for $\left\{ i,j,k,l\right\} =\left\{ 1,2,3,4\right\}$ we have $p_{i}+p_{j}=-p_{k}-p_{l}$ and therefore for all momenta on shell it holds
\begin{equation}
s_{ij}=2p_{i}\cdot p_{j}=2p_{k}\cdot p_{l}=-2p_{k}\cdot p_{ijk}\,.
\end{equation}
Here we have denoted $p_{ijk}=p_{i}+p_{j}+p_{k}$.
Let us define the following  semi-on-shell extension of the four-point amplitude with $p_i$, $p_j$ and $p_k$ on-shell and $p_{ijk}$ off-shell as
\begin{eqnarray}
\widetilde{A}_{4}\left( p_{i}^{a_{i}} ,p_{j}^{a_{j}} ,p_{k}^{a_{k}} ,-p_{ijk}^{a} \right)&=&-2\lambda _{a_{i}a_{j},a_{k}a}^{+}s_{ij}^{2} p_{k} \cdot p_{ijk} +\mathrm{cykl}\left( i,j,k\right) \notag\\
&&+\lambda _{a_{i}a_{j}a_{k}a}^{-}\left( s_{ij} -s_{ik} \right) \left( s_{ij} -s_{jk} \right)
\left( s_{ik} -s_{jk} \right).
\label{modified_4pt}
\end{eqnarray}
Inserting in $\widetilde{A}_{4}$ for $p_i$, $p_j$ and $p_k$ the deformed momenta which correspond to the six-point kinematics of the reconstructed 6pt amplitude,
we can  write for the factorization channel $\mathcal{F}=\{i,j,k\}$
\begin{equation}
\left. A_{4}\left( p_{i}^{a_{i}}(z) ,p_{j}^{a_{j}}(z) ,p_{k}^{a_{k}}(z) ,-p_{\cal{F}}^{i_{\cal{F}}}(z) \right)
\right\vert _{z=z_{\cal{F}}^{\pm}}=\left.\widetilde{A}_{4}\left( p_{i}^{a_{i}}(z) ,p_{j}^{a_{j}}(z) ,p_{k}^{a_{k}}(z) ,-p_{\cal{F}}^{i_{\cal{F}}}(z) \right)\right\vert _{z=z_{\cal{F}}^{\pm}}
\end{equation}
Then $\widetilde{A}_{4}( p_{i}^{a_{i}}(z) ,p_{j}^{a_{j}}(z) ,p_{k}^{a_{k}}(z) ,-p_{\cal{F}}^{i_{\cal{F}}}(z))$ can be considered as the desired modified four-point amplitude $\widetilde{A}_{4}\left( \mathcal{F}(z);i_{\cal{F}} \right)$ needed for the construction of the function $f_{\mathcal{F}}(z)$ in this factorization channel.
With this choice the physical meaning of the right-hand side of the formula (\ref{modified_6pt_recursion}) is clear.
The first term  has the explicit form\footnote{In this formula, the sum over factorization channels is identified with the sum over the sets of indices $\{i,j,k\}$ while the set $\{l,m,n\}$ corresponds to its complement. 
The overall factor $1/2$ is introduced in order to avoid double counting. }
\begin{eqnarray}
-\sum_{\mathcal{F}}f_{\mathcal{F}}(0)=-\frac{1}{2}\sum_{\{i,j,k\}
}\frac{\widetilde{A}_{4}( p_{i}^{a_{i}} ,p_{j}^{a_{j}} ,p_{k}^{a_{k}} ,-p_{ijk}^{a})\widetilde{A}_{4}( p_{l}^{a_{l}} ,p_{m}^{a_{m}} ,p_{n}^{a_{n}},p_{ijk}^{a})}{p_{ijk}^2},
\end{eqnarray}
and corresponds to the pole contributions with correct factorization properties in all factorization channels. 
The sum of the remaining two terms represents the contact contributions ensuring the right single soft limits.
Note that the pole terms are independent of $a_j$'s and therefore the possible dependence (if any) on the free parameters of the momentum deformation (\ref{deformation}) has to cancel within the contact terms.

Let us now concentrate on the second term on the right-hand side of (\ref{modified_6pt_recursion}).
Inserting the explicit form of the $\widetilde{A}_{4}\left( \mathcal{F}(z);i_{\cal{F}} \right)$ into the right-hand side of (\ref{modified_f}) we easily find, that for $\mathcal{F}=\{i,j,k\}$ and $\overline{\mathcal{F}}=\{1,2,3,4,5,6\}\backslash \mathcal{F}=\{l,m,n\}$ the function 
$f_{\mathcal{F}}(z) /z$ has only simple poles for $z=1/a_{r}$, $r=1,\dots,6$.
The residues are then  given after some algebra by the following formula
\begin{multline}
\mathrm{res}\left( \frac{f_{\mathcal{F}}(z) }{z},\frac{1}{a_{k}}\right) =\frac{a_{k}}{(a_{k}-a_{i})\left( a_{k}-a_{j}\right) }
s_{ij}\left\{ \lambda _{a_{i}a_{j},a_{k}a}^{+}\left[ s_{ik}\left(
a_{k}-a_{i}\right) +s_{jk}\left( a_{k}-a_{j}\right) \right] \right.\\
\left. -\lambda _{a_{i}a_{j}a_{k}a}^{-}\left[ s_{ik}\left(
a_{k}-a_{i}\right) -s_{jk}\left( a_{k}-a_{j}\right) \right] \right\} \\
\times \left\{ -\lambda _{a_{l}a_{m},a_{n}a}^{+}\frac{s_{lm}^{2}\left[
s_{ln}\left( a_{k}-a_{l}\right) +s_{mn}\left( a_{k}-a_{m}\right) \right] }{\left( a_{k}-a_{n}\right) }\right.\\
\left.+\frac{1}{3}\lambda _{a_{l}a_{m}a_{n}a}^{-}\frac{\left[
s_{ln}\left( a_{k}-a_{l}\right) -s_{mn}\left( a_{k}-a_{m}\right) \right] }{\left( a_{k}-a_{n}\right) }\right.\\
\left. \times \frac{\left[ s_{lm}\left( a_{k}-a_{l}\right) -s_{mn}\left(
a_{k}-a_{n}\right) \right] }{\left( a_{k}-a_{m}\right) }\frac{\left[
s_{lm}\left( a_{k}-a_{m}\right) -s_{ln}\left( a_{k}-a_{n}\right) \right] }{\left( a_{k}-a_{l}\right) }+\mathrm{cykl}\left( l,m,n\right) \right\}\,.
\end{multline}
After performing the summation over all the poles and all factorization channels we get finally the second term on the right-hand side of (\ref{modified_6pt_recursion}) in the form
\begin{multline}
-\sum_{\mathcal{F}}\sum_{l=1}^{6}\mathrm{res}\left( \frac{f_{\mathcal{F}%
}(z) }{z},\frac{1}{a_{l}}\right)
=-\frac{1}{4}\sum_{\sigma \in S_{6}}\frac{a_{\sigma \left( 3\right) }}{%
(a_{\sigma \left( 3\right) }-a_{\sigma \left( 1\right) })\left( a_{\sigma
\left( 3\right) }-a_{\sigma \left( 2\right) }\right) }s_{\sigma \left(
1\right) \sigma \left( 2\right) } \\
\times \left\{ \lambda _{a_{\sigma \left( 1\right) }a_{\sigma \left(
2\right) },a_{\sigma \left( 3\right) }a}^{+}\left[ s_{\sigma \left( 1\right)
\sigma \left( 3\right) }(a_{\sigma \left( 3\right) }-a_{\sigma \left(
1\right) })+s_{\sigma \left( 2\right) \sigma \left( 3\right) }\left(
a_{\sigma \left( 3\right) }-a_{\sigma \left( 2\right) }\right) \right]
\right. \\
\left. -\lambda _{a_{\sigma \left( 1\right) }a_{\sigma \left( 2\right)
}a_{\sigma \left( 3\right) }a}^{-}\left[ s_{\sigma \left( 1\right) \sigma
\left( 3\right) }(a_{\sigma \left( 3\right) }-a_{\sigma \left( 1\right)
})-s_{\sigma \left( 2\right) \sigma \left( 3\right) }\left( a_{\sigma \left(
3\right) }-a_{\sigma \left( 2\right) }\right) \right] \right\} \\
\times \left\{ -\lambda _{a_{\sigma \left( 4\right) }a_{\sigma \left(
5\right) },a_{\sigma \left( 6\right) }a}^{+}\frac{s_{\sigma \left( 4\right)
\sigma \left( 5\right) }^{2}\left[ s_{\sigma \left( 4\right) \sigma \left(
6\right) }\left( a_{\sigma \left( 3\right) }-a_{\sigma \left( 4\right)
}\right) +s_{\sigma \left( 5\right) \sigma \left( 6\right) }\left( a_{\sigma
\left( 3\right) }-a_{\sigma \left( 5\right) }\right) \right] }{\left(
a_{\sigma \left( 3\right) }-a_{\sigma \left( 6\right) }\right) }\right. \\
\left. +\frac{1}{3}\lambda _{a_{\sigma \left( 4\right) }a_{\sigma \left(
5\right) }a_{\sigma \left( 6\right) }a}^{-}\frac{\left[ s_{\sigma \left(
4\right) \sigma \left( 6\right) }\left( a_{\sigma \left( 3\right)
}-a_{\sigma \left( 4\right) }\right) -s_{\sigma \left( 5\right) \sigma
\left( 6\right) }\left( a_{\sigma \left( 3\right) }-a_{\sigma \left(
5\right) }\right) \right] }{\left( a_{\sigma \left( 3\right) }-a_{\sigma
\left( 6\right) }\right) }\right. \\
\left. \times \frac{\left[ s_{\sigma \left( 4\right) \sigma \left(
5\right) }\left( a_{\sigma \left( 3\right) }-a_{\sigma \left( 4\right)
}\right) -s_{\sigma \left( 5\right) \sigma \left( 6\right) }\left( a_{\sigma
\left( 3\right) }-a_{\sigma \left( 6\right) }\right) \right] }{\left(
a_{\sigma \left( 3\right) }-a_{\sigma \left( 5\right) }\right) }\right. \\
\left. \times \frac{\left[ s_{\sigma \left( 4\right) \sigma \left(
5\right) }\left( a_{\sigma \left( 3\right) }-a_{\sigma \left( 5\right)
}\right) -s_{\sigma \left( 4\right) \sigma \left( 6\right) }\left( a_{\sigma
\left( 3\right) }-a_{\sigma \left( 6\right) }\right) \right] }{\left(
a_{\sigma \left( 3\right) }-a_{\sigma \left( 4\right) }\right) }\right\}\,.
\label{residue_sum}
\end{multline}
Note that the result depends on the momentum deformation (\ref{deformation}) through the explicit dependence on $a_j$.
The same is true for the third term (\ref{third_term}) on the right-hand side of (\ref{modified_6pt_recursion}).
Provided the theory exists, the sum of these two terms have to be independent on the free parameters which parametrize the solution of the constraint (\ref{deformation_constraint}).
This necessary condition for the existence of the theory puts constraints on the 4pt  seed amplitudes. Note, however, that even when a solution of these constraints exists, the existence of the theory is not guaranteed.

\subsection{The bonus relation for the 6pt amplitude for $\sigma=2$}

Suppose now, that we require  the reconstructed six-point amplitude to obey the enhanced Adler zero with soft exponent $\sigma=2$.
In such a case, the third term on the right-hand side of (\ref{modified_6pt_recursion}) is missing and the only possible dependence on the parameters of the momentum shifts is encoded in (\ref{residue_sum}).
For the existence of the theory with the 4pt amplitudes (\ref{general_4pt}) and $O\left(
p^{2}\right) $ soft limit, it is therefore necessary that the above expression (\ref{residue_sum}) is
independent on the shift $a_{i}\rightarrow a_{i}+a$, $i=1,\ldots 6$ and on
the re-scaling $a_{i}\rightarrow \beta a_{i}$, $i=1,\ldots 6$. While the
latter condition is satisfied manifestly, the former is far from being
obvious. 
The necessary condition is obtained by shifting $a_{i}\rightarrow
a_{i}+a$, $i=1,\ldots 6$ and demanding the coefficient at $a$ to vanish:
\begin{eqnarray}
 &\phantom{=}&\sum_{\sigma \in S_{6}}\frac{s_{\sigma \left( 1\right) \sigma \left(
2\right) }}{(a_{\sigma \left( 3\right) }-a_{\sigma \left( 1\right) })\left(
a_{\sigma \left( 3\right) }-a_{\sigma \left( 2\right) }\right) \left(
a_{\sigma \left( 3\right) }-a_{\sigma \left( 4\right) }\right) \left(
a_{\sigma \left( 3\right) }-a_{\sigma \left( 5\right) }\right) \left(
a_{\sigma \left( 3\right) }-a_{\sigma \left( 6\right) }\right) }  \notag \\
&&\times \left\{ \lambda _{a_{\sigma \left( 1\right) }a_{\sigma \left(
2\right) },a_{\sigma \left( 3\right) }a}^{+}\left[ s_{\sigma \left( 1\right)
\sigma \left( 3\right) }(a_{\sigma \left( 3\right) }-a_{\sigma \left(
1\right) })+s_{\sigma \left( 2\right) \sigma \left( 3\right) }\left(
a_{\sigma \left( 3\right) }-a_{\sigma \left( 2\right) }\right) \right]
\right.  \notag \\
&&\left. -\lambda _{a_{\sigma \left( 1\right) }a_{\sigma \left( 2\right)
}a_{\sigma \left( 3\right) }a}^{-}\left[ s_{\sigma \left( 1\right) \sigma
\left( 3\right) }(a_{\sigma \left( 3\right) }-a_{\sigma \left( 1\right)
})-s_{\sigma \left( 2\right) \sigma \left( 3\right) }\left( a_{\sigma \left(
3\right) }-a_{\sigma \left( 2\right) }\right) \right] \right\}  \notag \\
&&\times \left\{ -\lambda _{a_{\sigma \left( 4\right) }a_{\sigma \left(
5\right) },a_{\sigma \left( 6\right) }a}^{+}\left[ s_{\sigma \left( 4\right)
\sigma \left( 6\right) }\left( a_{\sigma \left( 3\right) }-a_{\sigma \left(
4\right) }\right) +s_{\sigma \left( 5\right) \sigma \left( 6\right) }\left(
a_{\sigma \left( 3\right) }-a_{\sigma \left( 5\right) }\right) \right]
\right.  \notag \\
&&\times s_{\sigma \left( 4\right) \sigma \left( 5\right) }^{2}\left(
a_{\sigma \left( 3\right) }-a_{\sigma \left( 5\right) }\right) \left(
a_{\sigma \left( 3\right) }-a_{\sigma \left( 4\right) }\right)  \notag \\
&&\left. +\frac{1}{3}\lambda _{a_{\sigma \left( 4\right) }a_{\sigma \left(
5\right) }a_{\sigma \left( 6\right) }a}^{-}\left[ s_{\sigma \left( 4\right)
\sigma \left( 6\right) }\left( a_{\sigma \left( 3\right) }-a_{\sigma \left(
4\right) }\right) -s_{\sigma \left( 5\right) \sigma \left( 6\right) }\left(
a_{\sigma \left( 3\right) }-a_{\sigma \left( 5\right) }\right) \right]
\right.  \notag \\
&&\left. \times \left[ s_{\sigma \left( 4\right) \sigma \left( 5\right)
}\left( a_{\sigma \left( 3\right) }-a_{\sigma \left( 4\right) }\right)
-s_{\sigma \left( 5\right) \sigma \left( 6\right) }\left( a_{\sigma \left(
3\right) }-a_{\sigma \left( 6\right) }\right) \right] \right.  \notag \\
&&\left. \times \left[ s_{\sigma \left( 4\right) \sigma \left( 5\right)
}\left( a_{\sigma \left( 3\right) }-a_{\sigma \left( 5\right) }\right)
-s_{\sigma \left( 4\right) \sigma \left( 6\right) }\left( a_{\sigma \left(
3\right) }-a_{\sigma \left( 6\right) }\right) \right] \right\}=0\,.
\label{bonus_relation}
\end{eqnarray}
Note that this can be understood as the relation
\begin{equation}
\sum_{\mathcal{F}}\sum_{l=1}^{6}\mathrm{res}\left( f_{\mathcal{F}}(z) ,\frac{1}{a_{l}}\right) =0\,,
\end{equation}
which is a particular example of the bonus relation which has been mentioned in the introduction to this section. 
In general, bonus relations hold as a consequence of the better than necessary behavior of the function $f(z)$ for $z\to\infty$.
Indeed, in order to avoid  the unknown nonzero residue at infinity in the residue theorem (\ref{BCFW}), it suffices the $1/z$ fall off, while in fact in our case $f(z)=O(z^{-2})$  (c.f. (\ref{f_asymptotics})).
We can therefore apply the residue theorem to the function $f(z) $
instead of $f(z) /z$. 
Also in this case the residue at infinity
vanishes (c.f. (\ref{f_asymptotics})), and because there is no pole at $z=0$
we have only the unitarity poles\footnote{This is true under the condition  that we demand the enhanced Adler zero. 
In the general case, the bonus relations are modified due to the presence of the third term on the rhs of (\ref{modified_6pt_recursion}).}. 
Therefore
\begin{equation}
\sum_{\mathcal{F},I=\pm }\mathrm{res}\left( f(z) ,z_{\mathcal{F}%
}^{I}\right) =0\,.
\end{equation}
Repeating the above trick with replacing  $f(z)$ with $f_{\cal{F}}(z)$ and applying the residue theorem once more, we get
\begin{equation*}
\sum_{\mathcal{F},I=\pm }\mathrm{res}\left( f(z) ,z_{\mathcal{F}%
}^{I}\right) =\sum_{\mathcal{F}}\sum_{l=1}^{6}\mathrm{res}\left( f_{\mathcal{%
F}}(z) ,\frac{1}{a_{l}}\right).
\end{equation*}
The bonus relations in the form (\ref{bonus_relation}) are therefore 
necessary conditions for existence of the 6pt amplitude with demanded properties.
They give nontrivial constraints on the parameters of the four-point seed amplitudes and can be used for exploration of the landscape of the multi-flavor theories with Galileon power counting.

In the next section we will illustrate their applications on concrete examples.

\section{Analytical bootstrap  and examples of the classification of the theories}\label{applications_section}

In this section we will give explicit examples of the usefulness of the soft bootstrap techniques  introduced in the previous section for the classification of the multi-flavor theories with the Galileon power counting.
We will concentrate on two particular cases when the  relations which stem from the bootstrap methods can be solved analytically, namely on the classification of the  two-flavor theories with $\sigma=2$, and on the problem of the possible existence of the multi-flavor analogue of the Special Galileon.

\subsection{Bonus relations for two-flavor case with \texorpdfstring{$\sigma=2$}{sigma=2}}
\label{bonus2fl}
The parametrization of the 4pt amplitude simplifies in the case of two flavors since there is no totally antisymmetric tensor $\lambda _{ijkl}^{-}$  for $N=2$  and since, due to the symmetries (\ref{lambda_symmetries}) of $\lambda _{ijkl}^{+}$, we have only six independent couplings.
We get explicitly
\begin{eqnarray}
A_{4}\left( 1^{1},2^{1},3^{1},4^{1}\right) &=&\lambda _{1111}^{+}\left(
s_{12}^{3}+s_{13}^{3}+s_{23}^{3}\right) 
=3\lambda _{1111}^{+}s_{12}s_{13}s_{23} \notag\\
A_{4}\left( 1^{2},2^{2},3^{2},4^{2}\right) &=&\lambda _{2222}^{+}\left(
s_{12}^{3}+s_{13}^{3}+s_{23}^{3}\right) 
=3\lambda _{2222}^{+}s_{12}s_{13}s_{23} \notag\\
A_{4}\left( 1^{1},2^{1},3^{1},4^{2}\right) &=&\lambda _{1112}^{+}\left(
s_{12}^{3}+s_{13}^{3}+s_{23}^{3}\right) 
=3\lambda _{1112}^{+}s_{12}s_{13}s_{23}\notag \\
A_{4}\left( 1^{2},2^{2},3^{2},4^{1}\right) &=&\lambda _{2221}^{+}\left(
s_{12}^{3}+s_{13}^{3}+s_{23}^{3}\right) 
=3\lambda _{2221}^{+}s_{12}s_{13}s_{23} \notag\\
A_{4}\left( 1^{1},2^{1},3^{2},4^{2}\right) &=&\lambda
_{1122}^{+}s_{12}^{3}+\lambda _{1212}^{+}\left( s_{13}^{3}+s_{23}^{3}\right)\notag\\
&=&3\lambda
_{1212}^{+}s_{12}s_{13}s_{23}+\left( \lambda _{1122}^{+}-\lambda _{1212}^{+}\right) s_{12}^{3}.
\label{2_flavor_4pt}
\end{eqnarray}
Note that provided $\lambda _{1122}^{+}=\lambda _{1212}^{+}$,
the tensor $\lambda _{ijkl}^{+}$ is totally symmetric (cf. (\ref{lambda_symmetries})).
It is convenient to make use of the simplicity of the 4pt amplitudes and modify the general form of the bonus relation (\ref{bonus_relation}).
Namely, it is convenient,  in the defining formula for $f_{\mathcal{F}}(z)$
\begin{equation}
f_{\mathcal{F}}(z) =\sum_{b}\frac{\widetilde{A}_{4}\left( 
\mathcal{F}(z) \right) ^{b}\widetilde{A}_{4}\left( \overline{%
\mathcal{F}}(z) \right) ^{b}}{p_{\mathcal{F}}^{2}(z)
\prod\limits_{j=1}^{6}\left( 1-a_{j}z\right) ^{2}},
\end{equation}
to take the amplitudes $\widetilde{A}_{4}\left( \mathcal{F}(z) \right)
^{b}$ as
\begin{eqnarray}
\widetilde{A}_{4}\left( \left( \mathcal{F}(z) \right) \right)
^{b} =3\lambda _{aaab}^{+}s_{12}(z) s_{13}(z)
s_{23}(z) 
=3\lambda _{aaab}^{+}s_{12}s_{13}s_{23}\prod\limits_{i=1}^{3}\left(
1-a_{i}z\right) ^{2}
\label{4pt_aaab}
\end{eqnarray}
for $\mathcal{F=}\left\{ 1^{a},2^{a},3^{a}\right\} $ and similarly in other cases with the only exception of the amplitude $A_{4}\left( 1^{b},2^{a},3^{a},4^{b}\right)$, where we take
\begin{eqnarray}
\widetilde{A}_{4}\left( \left( \mathcal{F}(z) \right) \right)
^{b} &=&-2\left( \lambda _{aabb}^{+}-\lambda _{abab}^{+}\right)
s_{23}^{2}(z) p_{1}(z) \cdot p_{123}(z)
+3\lambda _{abab}^{+}s_{12}(z) s_{13}(z)
s_{23}(z)\notag \\
&=&-\left( \lambda _{aabb}^{+}-\lambda _{abab}^{+}\right) s_{23}^{2}\left[
s_{12}\left( 1-a_{2}z\right) +s_{13}\left( 1-a_{3}z\right) \right] \notag\\
&&\times \left( 1-a_{1}z\right) \left( 1-a_{2}z\right) ^{2}\left(
1-a_{3}z\right) ^{2} \notag\\
&&+3\lambda _{abab}^{+}s_{12}s_{13}s_{23}\left( 1-a_{1}z\right) ^{2}\left(
1-a_{2}z\right) ^{2}\left( 1-a_{3}z\right) ^{2}.
\label{4pt_baab}
\end{eqnarray}
Such a choice  minimizes the number of poles of
the function $f_{\mathcal{F}}(z)$ and simplifies
considerably the calculation of the left-hand sides of the bonus relations.
For instance, for the amplitude $A_{6}\left(
1^{a},2^{a},3^{a},4^{a},5^{a},6^{a}\right) $ where $a=1,2$, the function $f_{\mathcal{F}}(z)$ has only the unitarity poles and no $1/a_{i}$
poles, therefore the bonus relation is trivially satisfied.

Concerning the bonus relation for the amplitude $A_{6}\left(1^{b},2^{a},3^{a},4^{a},5^{a},6^{a}\right)$, where $a\ne b$, the factorization channel $\mathcal{F=}\left\{
1^{b},2^{a},3^{a}\right\}$ contributes only for internal flavor $b$.
Using (\ref{4pt_aaab}) and (\ref{4pt_baab}), we find that 
only the pole $z=1/a_{1}$ contributes and we get\footnote{Here and in the rest of this subsection, no summation over repeated flavor indices is assumed.}
\begin{eqnarray}
\mathrm{res}\left( f_{\mathcal{F}}(z) ,\frac{1}{a_{1}}\right)
&=&3\frac{\lambda _{aaab}^{+}s_{45}s_{46}s_{56}\left( \lambda
_{aabb}^{+}-\lambda _{abab}^{+}\right) s_{23}\left[ s_{12}\left(
a_{1}-a_{2}\right) +s_{13}\left( a_{1}-a_{3}\right) \right] }{\left(
a_{1}-a_{2}\right) \left( a_{1}-a_{3}\right) } \notag\\
&=&3s_{12}s_{13}s_{23}s_{45}s_{46}s_{56}\left[ \frac{1}{s_{12}\left(
a_{1}-a_{2}\right) }+\frac{1}{s_{13}\left( a_{1}-a_{3}\right) }\right] \notag\\
&&\times \left( \lambda _{aabb}^{+}-\lambda _{abab}^{+}\right) \lambda
_{aaab}^{+}\,.
\end{eqnarray}
For other factorization channels, the flavor structure is the same, while the labels \{2, 3, 4, 5, 6\}  are permuted. Therefore, the bonus relation reads (up to an overall combinatorial factor)
\begin{equation}
\left( \lambda _{aabb}^{+}-\lambda _{abab}^{+}\right) \lambda
_{aaab}^{+}K\left( s_{ij}\right) =0\,,
\end{equation}
where the kinematic factor is
\begin{eqnarray}
K\left( s_{ij}\right) &=&\sum_{\sigma \in S_{5}} \frac{s_{1\sigma \left( 3\right) }s_{\sigma \left( 2\right) \sigma \left(
3\right) }s_{\sigma \left( 4\right) \sigma \left( 5\right) }s_{\sigma \left(
4\right) \sigma \left( 6\right) }s_{\sigma \left( 5\right) \sigma \left(
6\right) }}{
\left( a_{1}-a_{\sigma \left( 2\right) }\right) } \,.
  \label{kin0}
\end{eqnarray}
Let us now concentrate on the bonus relation for amplitudes $A_{6}\left(1^{b},2^{b},3^{a},4^{a},5^{a},6^{a}\right)$, $a\ne b$.
There are two types of factorization channels, namely Type I ($\mathcal{F=}\left\{ 1^{b},2^{b},\sigma \left( 3\right) ^{a}\right\} $) and
Type II ($\mathcal{F=}\left\{ 1^{b},\sigma \left( 3\right) ^{a},\sigma \left( 4\right) ^{a}\right\} $) where $\sigma \in S_{4}$ is some permutation of $\left\{ 3,4,5,6\right\} $. The contribution of Type I (for simplicity let us take $\sigma =\mathrm{id}$, the pole of $f_{\mathcal{F}}(z) $ is then for $z=1/a_{3}$) has the following structure
\begin{eqnarray}
\mathrm{res}\left( f_{\mathcal{F}}(z) ,\frac{1}{a_{3}}\right)
&=&3s_{12}s_{13}s_{23}s_{45}s_{46}s_{56}\left[ \frac{1}{s_{23}\left(
a_{3}-a_{2}\right) }+\frac{1}{s_{13}\left( a_{3}-a_{1}\right) }\right] 
\notag\\&&\times\left( \lambda _{aabb}^{+}-\lambda _{abab}^{+}\right) \lambda
_{aaaa}^{+}\,.\label{type_I}
\end{eqnarray}
The contribution of Type II (again with $\sigma =\mathrm{id}$, now we get
poles for $z=1/a_{1}$ and $z=1/a_{2}$) reads
\begin{eqnarray}
&&\mathrm{res}\left( f_{\mathcal{F}}(z) ,\frac{1}{a_{1}}\right) +%
\mathrm{res}\left( f_{\mathcal{F}}(z) ,\frac{1}{a_{2}}\right)= \notag\\
&&=3\left( \lambda _{aabb}^{+}-\lambda _{abab}^{+}\right) \lambda
_{abab}^{+}s_{13}s_{14}s_{34}s_{25}s_{26}s_{56}  \notag\\
&&\times \left[ \frac{1}{s_{13}\left( a_{1}-a_{3}\right) }+\frac{1}{%
s_{14}\left( a_{1}-a_{4}\right) }+\frac{1}{s_{25}\left( a_{2}-a_{5}\right) }+%
\frac{1}{s_{26}\left( a_{2}-a_{6}\right) }\right]  \notag\\
&&+\left( \lambda _{aabb}^{+}-\lambda _{abab}^{+}\right) ^{2}\frac{%
s_{34}s_{56}}{\left( a_{1}-a_{2}\right) }\left\{ s_{56}s_{13}s_{14}\left[ 
\frac{1}{s_{14}\left( a_{1}-a_{4}\right) }+\frac{1}{s_{13}\left(
a_{1}-a_{3}\right) }\right] \right.  \notag\\
&&\left. \times \left[ s_{25}\left( a_{1}-a_{5}\right) +s_{26}\left(
a_{1}-a_{6}\right) \right] \right. \notag \\
&&\left. -s_{34}s_{25}s_{26}\left[ \frac{1}{s_{25}\left( a_{2}-a_{5}\right) }%
+\frac{1}{s_{26}\left( a_{2}-a_{6}\right) }\right] \left[ s_{13}\left(
a_{2}-a_{3}\right) +s_{14}\left( a_{2}-a_{4}\right) \right] \right\}\,.
\label{type_II}
\end{eqnarray}
Summing over all factorization channels we get finally the bonus relation in
the form
\begin{equation}
\left( \lambda _{aabb}^{+}-\lambda _{abab}^{+}\right) \left[\lambda
_{aaaa}^{+}K^{\left( I\right) }\left( s_{ij}\right)+\lambda
_{abab}^{+}K^{\left( II\right) }\left( s_{ij}\right)\right] +\left( \lambda
_{aabb}^{+}-\lambda _{abab}^{+}\right) ^{2}K^{\left( III\right) }\left(
s_{ij}\right) =0\,,
\label{bonus_relation_2}
\end{equation}
where the kinematical factors $K^{(I),(II)}(s_{ij})$ can be read off from (\ref{type_I}) and (\ref{type_II}) and their explicit form can be found in Appendix \ref{kinematical factors} (cf. (\ref{kin1}), (\ref{kin2})).

The last nontrivial bonus relation comes from the amplitudes $A_{6}\left(
1^{a},2^{a},3^{a},4^{b},5^{b},6^{b}\right) $, where $a\ne b$. The only type of the factorization channel which contributes
nontrivially is
\begin{equation}
\mathcal{F=}\left\{ \sigma \left( 1\right) ^{a},\sigma \left( 2\right)
^{a},\rho \left( 4\right) ^{b}\right\}\,,
\end{equation}
where $\sigma ,\rho \in S_{3}$ are permutations of $\left\{ 1,2,3\right\}$ and $\left\{ 4,5,6\right\} $ respectively. 
For $\sigma ,\rho =\mathrm{id}$ there are two contributions according to the flavor of the intermediate particle.
These have poles, namely for $z=1/a_{3}$ (for intermediate particle $a$) and for $z=1/a_{4}$ (for intermediate particle $b$). 
We get
\begin{eqnarray}
&&\mathrm{res}\left( f_{\mathcal{F}}(z) ,\frac{1}{a_{3}}\right) +%
\mathrm{res}\left( f_{\mathcal{F}}(z) ,\frac{1}{a_{4}}\right)= \notag\\
&&=3\left( \lambda _{aabb}^{+}-\lambda _{abab}^{+}\right) \lambda
_{aaab}^{+}s_{12}s_{14}s_{24}s_{35}s_{36}s_{56} \left[ \frac{1}{s_{35}\left( a_{3}-a_{5}\right) }+\frac{1}{%
s_{36}\left( a_{3}-a_{6}\right) }\right] \notag\\
&&+\,\,3\left( \lambda _{aabb}^{+}-\lambda _{abab}^{+}\right) \lambda
_{bbba}^{+}s_{12}s_{14}s_{24}s_{35}s_{36}s_{56} \left[ \frac{1}{s_{14}\left( a_{4}-a_{5}\right) }+\frac{1}{%
s_{24}\left( a_{4}-a_{6}\right) }\right],
\end{eqnarray}
Summing up the contributions of all the factorization channels we get the final form of the corresponding bonus relation  (up to a combinatoric factor) in the form
\begin{equation}
\left( \lambda _{aabb}^{+}-\lambda _{abab}^{+}\right) \lambda
_{aaab}^{+}K^{\left( IV\right) }\left( s_{ij}\right) +\left( \lambda
_{aabb}^{+}-\lambda _{abab}^{+}\right) \lambda _{bbba}^{+}K^{\left(
V\right) }\left( s_{ij}\right) =0,
\label{bonus_relation_3}
\end{equation}
with the  kinematical  factors $K^{(IV),(V)}(s_{ij})$. 
These are listed in Appendix \ref{kinematical factors} (cf. (\ref{kin3}), (\ref{kin4})). 

To summarize, we have found the following nontrivial bonus relations
\begin{eqnarray}
&&\left( \lambda _{aabb}^{+}-\lambda _{abab}^{+}\right) \lambda
_{aaab}^{+}K\left( s_{ij}\right) =0, \notag\\
&&\left( \lambda _{aabb}^{+}-\lambda _{abab}^{+}\right) \left[\lambda
_{aaaa}^{+}K^{\left( I\right) }\left( s_{ij}\right)+\lambda
_{abab}^{+}K^{\left( II\right) }\left( s_{ij}\right) +\left( \lambda
_{aabb}^{+}-\lambda _{abab}^{+}\right)K^{\left( III\right) }\left(
s_{ij}\right)\right] =0,\notag\\
&&\left( \lambda _{aabb}^{+}-\lambda _{abab}^{+}\right) \lambda
_{aaab}^{+}K^{\left( IV\right) }\left( s_{ij}\right) +\left( \lambda
_{aabb}^{+}-\lambda _{abab}^{+}\right) \lambda _{bbba}^{+}K^{\left(
V\right) }\left( s_{ij}\right) =0\,,
\label{2_flavor_bonus_relations}
\end{eqnarray}
where the kinematic factors $K,K^{(I)},\ldots ,K^{\left( V\right) }$ are known functions of the invariants $s_{ij}$. 
The dependence on $s_{ij}$ is both explicit and implicit, in the latter case  through the differences $a_i-a_j$.
It is given explicitly by (\ref{kin0}), (\ref{kin1}), (\ref{kin2}), (\ref{kin3}) and (\ref{kin4}). 
Note that the differences $a_{i}-a_{j}$ are determined uniquely up to an overall normalization (see (\ref{a_difference})). 
Since the kinematic factors $K,K^{(I)},\ldots ,K^{\left( V\right) }$ are homogeneous function of $a_{i}-a_{j}$ with the same degree, this overall normalization is irrelevant.
Up to this  irrelevant overall normalization factor, the bonus relations are invariant with respect to the change of the parameters which parametrize the solution of the constraint (\ref{deformation_constraint}). 

The sufficient condition for the validity of the bonus relations is  $\lambda _{aabb}^{+}=\lambda _{abab}^{+}$.
This condition means that the
tensor $\lambda _{ijkl}$ is totally symmetric. 
This corresponds to the two-flavor multi-Galileon amplitudes (\ref{contact_4pt}) and therefore to the multi-Galileon theory without cubic vertices (or its dual).

On the other hand, provided $K\neq 0$ for generic configuration (this assumption was tested and proved numerically), we get
\begin{equation}
\left( \lambda _{aabb}^{+}-\lambda _{abab}^{+}\right) \lambda _{aaab}^{+}=0\,,
\end{equation}
and the third bonus relation is then trivially satisfied. 
If $\lambda_{aaab}^{+}\neq 0$, we get  $\lambda _{aabb}^{+}=\lambda _{abab}^{+}$ 
and also the second bonus relation is trivial. 
Assuming $\lambda _{aaab}^{+}=0$, the second bonus relations gives either $\lambda _{aabb}^{+}=\lambda _{abab}^{+}$, or
\begin{equation}
  \lambda
_{aaaa}^{+}K^{\left( I\right) }\left( s_{ij}\right)+\lambda
_{abab}^{+}K^{\left( II\right) }\left( s_{ij}\right) +\left( \lambda
_{aabb}^{+}-\lambda _{abab}^{+}\right)K^{\left( III\right) }\left(
s_{ij}\right)=0.  
\end{equation}
In the latter case, provided $K^{\left( I\right) }$, $K^{\left( II\right) }$ and $K^{\left( III\right) }$ are linearly independent functions (these assumptions we tested and proved numerically), we get
\begin{equation}
    \lambda_{aaaa}^{+}= \lambda_{abab}^{+}=\lambda _{aabb}^{+}=0
\end{equation}
and the 4pt amplitudes are trivial.
Therefore, for nontrivial 4pt amplitudes, the equality $\lambda_{aabb}^{+}=\lambda _{abab}^{+}$ is also a necessary condition for the validity of the bonus relations.

To summarize,  the bonus relations (\ref{2_flavor_bonus_relations}), which are the necessary conditions for the existence of a two-flavor theory with Galileon power counting and $O\left( p^{2}\right) $ soft limit in $D=4$, constrain the 4pt amplitudes to be the amplitudes  derived form the Lagrangian of the form  (\ref{Lagrangian_basic}) without cubic vertices, or its dual in the sense of Section \ref{section_duality}.
We will return back to this statement for a general number of flavors in Section~\ref{sec:7}.

\subsection{Soft bootstrap  for multi-flavor Special Galileon}

In this subsection we will discuss the possible multi-flavor generalization of the Special Galileon, which in the single-flavor case behaves even better in the single soft limit, namely as $O(p^3)$.
In close analogy with the single-flavor case, we will seek for such a theory within the subclass of theories with vanishing five-point seed amplitudes and with the seed four-point amplitudes with totally
symmetric constants $\lambda _{ijkl}^{+}$, namely
\begin{equation}
A_{4}\left( 1^{a_{1}},2^{a_{2}},3^{a_{3}},4^{a_{4}}\right) =\lambda
_{a_{1}a_{2}a_{3}a_{4}}^{+}s_{12}s_{13}s_{23}.  \label{seed_sGal}
\end{equation}
Assume now, that the 6pt amplitudes have $O\left( p^{3}\right) $ single soft limit.
Then the function
\begin{equation}
f(z) =\frac{A_{6}\left( p_{1}^{a_{1}}(z) ,\ldots
,p_{6}^{a_{6}}(z) \right) }{\prod\limits_{j=1}^{6}\left(
1-a_{j}z\right) ^{3}}\overset{z\rightarrow \infty }{=}O\left( z^{-8}\right)
\end{equation}
has only the unitarity poles.
This function can be then used for the   soft BCFW recursion which can be rewritten in the form  (\ref{modified_6pt_recursion}), now without the last term.
Therefore, the six point amplitude can be expressed as
\begin{equation}
A_{6}\left( p_{1},\ldots ,p_{n}\right) =-\sum_{\mathcal{F}}f_{\mathcal{F}}(0) -\sum_{\mathcal{F}}\sum_{j=1}^{6}\mathrm{res}\left( 
\frac{f_{\mathcal{F}}(z) }{z},\frac{1}{a_{j}}\right)\,,
\label{amplitude_special_multi_galileon}
\end{equation}
where for given factorization channel, e.g. $\mathcal{F=}\left\{1^{a_{1}},2^{a_{2}},3^{a_{3}}\right\}$, we define
\begin{eqnarray}
f_{\mathcal{F}}(z) &=&\lambda _{a_{1}a_{2}a_{3}a}^{+}\lambda
_{a_{4}a_{5}a_{6}a}^{+}\frac{s_{12}(z) s_{13}(z)
s_{23}(z) s_{45}(z) s_{46}(z)
s_{56}(z) }{p_{\mathcal{F}}^{2}(z)
\prod\limits_{j=1}^{6}\left( 1-a_{j}z\right) ^{3}} \notag\\
&=&\lambda _{a_{1}a_{2}a_{3}a}^{+}\lambda _{a_{4}a_{5}a_{6}a}^{+}\frac{%
s_{12}s_{13}s_{23}s_{45}s_{46}s_{56}}{p_{\mathcal{F}}^{2}(z)
\prod\limits_{j=1}^{6}\left( 1-a_{j}z\right) }\,.
\end{eqnarray}
For the individual terms on the right hand side of (\ref{amplitude_special_multi_galileon}) we get explicitly
\begin{eqnarray}
\sum_{\mathcal{F}}f_{\mathcal{F}%
}(0)&=&\left( \frac{1}{3!}\right) ^{2}\sum_{\sigma \in S_{6}}\lambda _{a_{\sigma
\left( 1\right) }a_{\sigma \left( 2\right) }a_{\sigma \left( 3\right)
}a}^{+}\lambda _{a_{\sigma \left( 4\right) }a_{\sigma \left( 5\right)
}a_{\sigma \left( 6\right) }a}^{+} \notag\\
&&\times
\frac{s_{\sigma \left( 1\right) \sigma \left( 2\right) }s_{\sigma
\left( 1\right) \sigma \left( 3\right) }s_{\sigma \left( 2\right) \sigma
\left( 3\right) }s_{\sigma \left( 4\right) \sigma \left( 5\right) }s_{\sigma
\left( 4\right) \sigma \left( 6\right) }s_{\sigma \left( 5\right) \sigma
\left( 6\right) }}{P_{\sigma \left( 1\right) \sigma \left( 2\right) \sigma
\left( 3\right) }^{2}}, 
\label{pole_term}
\end{eqnarray}
and
\begin{multline}
\sum_{\mathcal{F}}\sum_{j=1}^{6}\mathrm{res}\left( \frac{%
f_{\mathcal{F}}(z) }{z},\frac{1}{a_{j}}\right)= \frac{1}{3!}\sum_{\sigma \in S_{6}}\lambda _{a_{\sigma
\left( 1\right) }a_{\sigma \left( 2\right) }a_{\sigma \left( 3\right)
}a}^{+}\lambda _{a_{\sigma \left( 4\right) }a_{\sigma \left( 5\right)
}a_{\sigma \left( 6\right) }a}^{+}a_{\sigma \left( 1\right) }^{9}\\
\times \frac{s_{\sigma \left( 1\right) \sigma \left( 2\right) }s_{\sigma
\left( 1\right) \sigma \left( 3\right) }s_{\sigma \left( 4\right) \sigma \left( 5\right) }s_{\sigma
\left( 4\right) \sigma \left( 6\right) }s_{\sigma \left( 5\right) \sigma
\left( 6\right) }}{\left( a_{\sigma \left( 1\right) }-a_{\sigma \left(
2\right) }\right) \left( a_{\sigma \left( 1\right) }-a_{\sigma \left(
3\right) }\right)\prod\limits_{j\neq \sigma \left( 1\right) }\left(
a_{\sigma \left( 1\right) }-a_{j}\right) }\,. \label{counterterm}
\end{multline}
The first term on the right-hand side of (\ref{amplitude_special_multi_galileon}) corresponds to the pole terms with right factorization while
the second term should be then a contact counterterm, which ensures the soft limit. 

The  resulting amplitude (\ref{amplitude_special_multi_galileon}) as a whole has to be independent on the choice of $a_{i}$'s. 
It is easy to see that the independence on the shift $a_{i}\rightarrow a_{i}+a$ is equivalent to
the  bonus relations (here $k=0,\dots,8$)
\begin{eqnarray}
 &&\sum_{\sigma \in S_{6}}\lambda _{a_{\sigma \left( 1\right) }a_{\sigma
\left( 2\right) }a_{\sigma \left( 3\right) }a}^{+}\lambda _{a_{\sigma \left(
4\right) }a_{\sigma \left( 5\right) }a_{\sigma \left( 6\right) }a}^{+}{s_{\sigma \left( 1\right) \sigma \left( 2\right) }s_{\sigma
\left( 1\right) \sigma \left( 3\right) }s_{\sigma \left( 4\right) \sigma \left( 5\right) }s_{\sigma
\left( 4\right) \sigma \left( 6\right) }s_{\sigma \left( 5\right) \sigma
\left( 6\right) }}
  \notag \\
&&\times \frac{a_{\sigma \left( 1\right) }^{k}}{\left( a_{\sigma \left( 1\right) }-a_{\sigma \left( 2\right)
}\right) \left( a_{\sigma \left( 1\right) }-a_{\sigma \left( 3\right)}\right)\prod\limits_{j\neq \sigma \left( 1\right) }\left(
a_{\sigma \left( 1\right) }-a_{j}\right) }=0\,,\label{bonus_sGal}
\end{eqnarray}
which are discussed in detail in Appendix \ref{appendix_bonus_relations},
while the requirement of the independence on the scaling $a_{i}\rightarrow \alpha \alpha _{i}$ implies, that the right-hand side of (\ref{counterterm}) vanishes identically, %
since this term scales as $\alpha ^{2}$ when $a_j\to\alpha a_j$. 
The amplitude  is then necessarily reduced to the sum of the pole terms
\begin{eqnarray}
A_{6}\left( p_{1},\ldots ,p_{n}\right) &=&\left( \frac{1}{3!}\right)
^{2}\sum_{\sigma \in S_{6}}\lambda _{a_{\sigma \left( 1\right) }a_{\sigma
\left( 2\right) }a_{\sigma \left( 3\right) }a}^{+}\lambda _{a_{\sigma \left(
4\right) }a_{\sigma \left( 5\right) }a_{\sigma \left( 6\right) }a}^{+} 
\notag \\
&&\times \frac{s_{\sigma \left( 1\right) \sigma \left( 2\right) }s_{\sigma
\left( 1\right) \sigma \left( 3\right) }s_{\sigma \left( 2\right) \sigma
\left( 3\right) }s_{\sigma \left( 4\right) \sigma \left( 5\right) }s_{\sigma
\left( 4\right) \sigma \left( 6\right) }s_{\sigma \left( 5\right) \sigma
\left( 6\right) }}{P_{\sigma \left( 1\right) \sigma \left( 2\right) \sigma
\left( 3\right) }^{2}}.  \label{amplitude_sGal}
\end{eqnarray}
On the other hand,  such an amplitude does not obey the desired soft behavior automatically for general $\lambda^{+}_{a_1 a_2 a_3 a_4}$.
This can be used in order to find an additional necessary condition.
In the soft kinematics 
\begin{equation}
    p_1\to t p_1,\quad p_i\to p_i(t), \quad p_i(0)=p_i,\quad i=2,\dots,6,\quad \sum_{i=2}^{6}p_i=0\,,
\end{equation}
described in Section \ref{soft_theorem_section}, and in the soft limit  $t\rightarrow 0$ we get from (\ref{amplitude_sGal})
\begin{eqnarray}
A_{6}\left(t 1^{a_{1}},2(t)^{a_{2}}\ldots ,6(t)^{a_{6}}\right)&& \overset{t\rightarrow 0}%
{=}\frac{t^2}{2!3!}\sum_{\sigma \in S_{5}}\lambda _{a_{1}a_{\sigma \left(
2\right) }a_{\sigma \left( 3\right) }a}^{+}\lambda _{a_{\sigma \left(
4\right) }a_{\sigma \left( 5\right) }a_{\sigma \left( 6\right) }a}^{+} 
\notag \\
&&\times s_{1\sigma \left( 2\right) }s_{1\sigma \left( 3\right) }s_{\sigma
\left( 4\right) \sigma \left( 5\right) }s_{\sigma \left( 4\right) \sigma
\left( 6\right) }s_{\sigma \left( 5\right) \sigma \left( 6\right) }+O(t^3)\,, 
\end{eqnarray}
where we sum over all permutations $\sigma\in S_5$ of the indices $\{2,3,4,5,6\}$.
Therefore, the \emph{necessary} condition for the $O\left(t^{3}\right) $
enhanced soft limit reads
\begin{equation}
\sum_{\sigma \in S_{5}}\lambda _{a_{1}a_{\sigma \left( 2\right) }a_{\sigma
\left( 3\right) }a}^{+}\lambda _{a_{\sigma \left( 4\right) }a_{\sigma \left(
5\right) }a_{\sigma \left( 6\right) }a}^{+}s_{1\sigma \left( 2\right)
}s_{1\sigma \left( 3\right) }s_{\sigma \left( 4\right) \sigma \left(
5\right) }s_{\sigma \left( 4\right) \sigma \left( 6\right) }s_{\sigma \left(
5\right) \sigma \left( 6\right) }=0
 \label{multi_s_Gal_soft_limit}
\end{equation}
for any configuration corresponding to the  kinematic  
\begin{equation}
p_{i}^{2} =0,~~~~i=1,2,\ldots ,6,  \,\,\,\,
\sum_{i=2}^{6}p_{i} =0.  \label{soft_kinematics_1}
\end{equation}
The condition (\ref{multi_s_Gal_soft_limit}) can be easily understood in the \emph{single} Galileon case, when there is only one four-point coupling constant, i.e.  $\lambda_{a_1 a_2 a_3 a_4}^{+}\to \lambda_4$.
Indeed, we get for the soft kinematics (\ref{soft_kinematics_1})\footnote{This can be proved by direct calculation using the independent $s_{ij}$'s, i.e. using the soft kinematic to eliminate $p_{6}=-p_{2}-p_{3}-p_{4}-p_{5}$ (cf. (\ref{soft_kinematics_1}))
and one of the $s_{ij}$, $i=2,3,4,5$ using the constraint $p_{6}^{2}=0$.}
\begin{equation}
\sum_{\sigma \in S_{5}}s_{1\sigma \left( 2\right) }s_{1\sigma \left(
3\right) }s_{\sigma \left( 4\right) \sigma \left( 5\right) }s_{\sigma \left(
4\right) \sigma \left( 6\right) }s_{\sigma \left( 5\right) \sigma \left(
6\right) }=192\ G( 1,2,3,4,5)\,,  
\label{s_Gal_soft_limit}
\end{equation}
where $G(1,2,3,4,5)$ is the Gram determinant. 
Therefore the right-hand side of (\ref{s_Gal_soft_limit}) vanishes in $D=4$ automatically in the single Galileon case\footnote{ In $D>4$, there is a counterterm available  which can be added to the amplitude and which stems from the 6pt special Galileon Lagrangian.}.

Due to the symmetry of the couplings $\lambda_{a_{1}a_{2}a_{3}a_{4}}^{+}$, there are in fact only ten different terms on the right
hand side of (\ref{multi_s_Gal_soft_limit}), which are in one-to-one correspondence with the different Feynman-graph-like contributions to the amplitude $A_6$ (cf. (\ref{amplitude_sGal})).
After imposing the soft kinematic (\ref{soft_kinematics_1}) to eliminate the redundant Mandelstams in (\ref{multi_s_Gal_soft_limit}), namely using the constraints 
\begin{eqnarray}
s_{i6}&=&-\sum_{j=2}^{5}s_{ij},\notag\\s_{23}&+&s_{24}+s_{25}+s_{34}+s_{35}+s_{45}=0\,,
\end{eqnarray}
the right-hand side becomes a linear combination of monomials of the fifth order in the nine Mandelstam variables. 
In $D>4$, the above fifth order monomials are linearly independent, while for $D=4$, there is one \emph{and only one} linear combination of these monomials (up to an overall factor) which vanishes, namely the Gram determinant  $G(1,2,3,4,5)$. 
Note that for the soft kinematics, the relation (\ref{s_Gal_soft_limit}) holds. 
Therefore in order to satisfy (\ref{multi_s_Gal_soft_limit}) in $D=4$, the left-hand side has to be a multiple of $G(1,2,3,4,5)$ which is possible only if
\begin{equation}
\lambda _{a_{1}a_{\sigma \left( 2\right) }a_{\sigma \left( 3\right)
}a}^{+}\lambda _{a_{\sigma \left( 4\right) }a_{\sigma \left( 5\right)
}a_{\sigma \left( 6\right) }a}^{+}=\lambda
_{a_{1}a_{2}a_{3}a}^{+}\lambda _{a_{4}a_{5}a_{6}a}^{+}=\text{const}\,.
\label{necessary_and_sufficient_for_sGal}
\end{equation}
This is the desired necessary condition.
In fact, it is also a sufficient condition for the validity of the bonus relations (\ref{bonus_sGal}) and the scaling independence relation.
Indeed, in the case of single Galileon, these conditions must be satisfied automatically, since the theory exists. This means for $n=0,1,\ldots ,9$ 
\begin{eqnarray}
 &&\sum_{\sigma \in S_{6}}a_{\sigma \left( 1\right) }^{n}
 \frac{s_{\sigma
\left( 1\right) \sigma \left( 3\right) }s_{\sigma \left( 4\right) \sigma \left( 5\right) }s_{\sigma
\left( 4\right) \sigma \left( 6\right) }s_{\sigma \left( 5\right) \sigma
\left( 6\right) }}{\left( a_{\sigma \left( 1\right) }-a_{\sigma \left( 2\right)
}\right) \left( a_{\sigma \left( 1\right) }-a_{\sigma \left( 3\right)}\right)\prod\limits_{j\neq \sigma \left( 1\right) }\left(
a_{\sigma \left( 1\right) }-a_{j}\right) }=0.  \label{bonus_single_sGal}
\end{eqnarray}
Therefore, provided (\ref{necessary_and_sufficient_for_sGal}) holds, we can factor out the constant $\lambda
_{a_{1}a_{2}a_{3}a}^{+}\lambda _{a_{4}a_{5}a_{6}a}^{+}$ in the bonus relations (\ref{bonus_sGal}) and the vanishing of the right-hand sides of (\ref{counterterm}) are trivially satisfied.

\subsubsection{Solution of the necessary and sufficient conditions  \label{Solution}}

In this subsection we will prove that  the multi-flavor analogue of the Special Galileon, as we have defined it in the introduction to this section,  is up to an $SO(N)$ rotation equivalent to the sum of $N$ independent single Special Galileons.

Note that due to the total symmetry of the tensor $\lambda _{ijkl}^{+}$, the necessary and sufficient condition (\ref{necessary_and_sufficient_for_sGal}) means
\begin{equation}
\lambda _{a_{1}a_{2}a_{3}a}^{+}\lambda _{a_{4}a_{5}a_{6}a}^{+}=\lambda
_{a_{4}a_{2}a_{3}a}^{+}\lambda _{a_{1}a_{5}a_{6}a}^{+}\,.
\label{lambda4_simplified}
\end{equation}
This is valid e.g. when $a_{1}=a_{2}=\ldots =a_{6}$ or when all but one index $a_{i}$ are the same.
The solution of the general case is similar as in Subsection \ref{solution_of_necessary_condition}. Defining a symmetric
matrices $\Lambda^{(cd) }$ as
\begin{equation}
\Lambda_{ab}^{(cd)} = \lambda_{acdb}^{+}\,,
\end{equation}
we can rewrite (\ref{lambda4_simplified}) in the form 
\begin{equation}
\bigl[ \Lambda ^{(ab)}, \Lambda^{(cd)}\bigr] =0\,.
\end{equation}
It is therefore possible to diagonalize the matrices $\Lambda ^{(cd) }$
simultaneously by orthogonal matrix $M$, (here no summation over $m$ is understood)
\begin{equation}
M_{me}M_{nf}\Lambda _{ef}^{(ab) }=\alpha _{m}^{(ab)
}\delta _{mn},
\end{equation}
and defining
\begin{equation}
M_{ac}M_{bd}\alpha _{m}^{(cd) }\equiv\beta _{m}^{(ab) },
\end{equation}
it is possible by field rotation to convert the general case to
\begin{equation}
\lambda _{mabn}^{+}\rightarrow M_{ac}M_{bd}M_{me}M_{nf}\Lambda _{ef}^{(cd) }=\beta _{m}^{(ab) }\delta _{mn}.
\end{equation}
But the rotated $\lambda _{mabn}^{+}$ must be totally symmetric, therefore
\begin{equation}
\beta _{m}^{(ab) }\delta _{mn}=\beta _{m}^{(an)}\delta _{mb}=\beta _{m}^{(nb)}\delta _{ma}=\beta _{b}^{(mn)}\delta _{ab}\,.
\end{equation}
As a consequence,  for $m=n$ we get
\begin{equation}
\beta _{m}^{(ab) }=\beta _{m}^{(am) }\delta_{mb}=\beta _{m}^{(mb) }\delta _{ma}=\beta _{b}^{(mm) }\delta _{ab}\,,
\end{equation}
and therefore
\begin{equation}
\beta _{m}^{(ab) }=\beta _{m}^{(mm) }\delta
_{mb}\delta _{ma}.
\end{equation}
Finally 
\begin{equation*}
\lambda _{mabn}^{+}=\beta _{m}^{(mm)}\delta _{mn}\delta_{mb}\delta _{ma}
\end{equation*}
and therefore in the rotated field basis we get the seed 4pt amplitudes decoupled and corresponding to the theory which is a sum of independent Special Galileons.
Therefore, according to the  soft BCFW recursion, all the amplitudes behave as $O(p^3)$ in the soft limit and the theory is equivalent after an appropriate $SO(N)$ rotation to the $N$ exemplars of the single-flavor Special Galileon.

\subsubsection{Note on the Lagrangian approach}

Of course, some of the above results can also be obtained directly from the Lagrangian approach. 
Suppose we try to construct  the multi-flavor Special Galileon by means of defining its Lagrangian in analogy with the single-Galileon case in $D=4$ assuming that it is set by (\ref{Lagrangian_basic}) with  $\mathcal{L}_{3}=\mathcal{L}_{5}=0$\footnote{This implies that $\lambda _{i_{1}i_{2}i_{3} i_{4}}^{+}$ is totally symmetric and there are no 5pt (seed) amplitudes, in accordance with the assumptions of the amplitude approach.} and 
\begin{equation}
\mathcal{L}_{4}=-\frac{1}{3!}\lambda _{i_{1}i_{2}i_{3} i_{4}}^{+}\phi
_{i_{4}}\delta _{\nu _{1}\ldots \nu _{3}}^{\mu _{1}\ldots \mu
_{3}}\prod\limits_{j=1}^{3}\partial _{\mu _{j}}\partial ^{\nu _{j}}\phi
_{i_{j}}.
\end{equation}
Then the 4pt Feynman rule 
\begin{equation}
V_{4}^{i_{1}\ldots i_{4}}=4\lambda _{i_{1}\ldots i_{4}}^{+}G(1,2,3) =\lambda _{i_{1}\ldots i_{4}}^{+}s_{12}s_{13}s_{23}
\end{equation}
reproduces the amplitude (\ref{seed_sGal}) and the 6pt amplitude is just the pure pole part corresponding to  eq. (\ref{amplitude_sGal}), in accordance with the results of the amplitude approach discussed in the previous subsection.

Therefore we can repeat the discussion  following after  eq. (\ref{amplitude_sGal}) and conclude, that the necessary and sufficient condition  for $O\left( p^{3}\right) $ single soft
limit of the 6pt amplitude $A_{6}\left( 1^{a_{1}},\ldots ,6^{a_{6}}\right) $ is again the relation (\ref{necessary_and_sufficient_for_sGal}).
According to subsection \ref{Solution}, this implies that within the class of the  the theories in $D=4$ with Lagrangian  (\ref{Lagrangian_basic}) and with $\mathcal{L}_{3}=\mathcal{L}_{5}=0$ the only theory with enhanced $O(p^3)$ soft limit is up
to the $SO(N)$ rotation equivalent to the sum of $N$ independent single Special Galileons.

\section{On-shell reconstruction and numerical bootstrap}\label{sec:7}

In the previous section we discussed two simple examples of the application of the bonus relation to the classification of the multi-flavor theories with Galileon power counting.
The common feature of these examples was a relatively  small number of the independent constraints, stemming either from the bonus relation or from the consistency relations, as well as the possibility to solve  them  analytically.
However, with an increasing number of flavors, the number of independent relations grows rapidly and  this feature obscures the attempts to find their solution analytically either for generic $N$ or for particular fixed $N>2$.
The best we can do in this cases is to fix the number of flavors and try to explore the validity of the relations numerically.
This is done by means of generating an appropriate set of kinematic configurations randomly, inserting them into the  relations and solving the constraints for the couplings numerically.

In this section we will formulate a conjecture which generalizes the analytical results concerning the classification of the two-flavor case  to the general case of  multi-flavor theory with Galileon power counting and present some results of the numerical tests of this conjecture using the  approach described above and applied for concrete $N$'s.
In what follows we will be working exclusively in $D=4$.

\subsection{Numerical bootstrap for theories with enhanced soft limit}
As we have discussed in the previous sections, the bottom-up method of constructing the tree-level amplitudes via recursion relations depends on the explicit construction of the set of all possible  seed amplitudes. 
In our case for the Galileon-like theories in $D=4$ this set includes the 4-pt and 5-pt contact terms, with six and eight derivatives, respectively. 
The BCFW soft recursive formula then fixes the form of all other higher point tree-level amplitudes. 
However, as already discussed, not all sets of seed amplitudes will be the healthy ones. 
We have to check the 4pt and 5pt input for the self-consistency. 
For the 5pt vertices (equal to the 5pt amplitudes) we can immediately check their soft behaviour. Unfortunately, this is not an option for the 4pt vertices as the kinematics is too restrictive. 
The consistency check can be done, however, employing the  bonus relations introduced in Sec.\ref{bonus2fl} which can be easily implemented for more flavors systematically via a numerical algorithm. 
The outcome of this analysis up to the  five-flavor cases will be briefly described here.

We have already discussed the parametrization of the seed 4pt amplitudes of the theories with Galileon power counting  in Sec.~\ref{sec:rec6}.
Let us note that the resulting formula (\ref{general_4pt})  can be also written in the form, which has  manifest
the totally symmetric term, which corresponds to the Lagrangian (\ref{Lagrangian_basic}) of the
multi-Galileon without cubic interactions, and the remaining ``offending staff'' 
\begin{eqnarray}
A_{4}\left( 1^{a_{1}},2^{a_{2}},3^{a_{3}},4^{a_{4}}\right) &=&3\lambda
_{a_{1}a_{2}a_{3}a_{4}}^{0}s_{12}s_{13}s_{23} +\widetilde{\lambda }_{a_{1}a_{2},a_{3}a_{4}}^{+}s_{12}^{3}+\widetilde{%
\lambda }_{a_{1}a_{3},a_{2}a_{4}}^{+}s_{13}^{3}+\widetilde{\lambda }%
_{a_{2}a_{3},a_{1}a_{4}}^{+}s_{23}^{3}\notag \\
&&+\lambda _{a_{1}a_{2}a_{3}a_{4}}^{-}\left( s_{12}-s_{13}\right) \left(
s_{12}-s_{23}\right) \left( s_{13}-s_{23}\right),
\label{4pt_decomposition}
\end{eqnarray}%
where the new couplings are%
\begin{eqnarray}
\lambda _{a_{1}a_{2}a_{3}a_{4}}^{0} &=&\frac{1}{3}\left( \lambda
_{a_{1}a_{2},a_{3}a_{4}}^{+}+\lambda _{a_{1}a_{3},a_{2}a_{4}}^{+}+\lambda
_{a_{2}a_{3},a_{1}a_{4}}^{+}\right) \notag\\
\widetilde{\lambda }_{a_{1}a_{2},a_{3}a_{4}}^{+} &=&\lambda
_{a_{1}a_{2},a_{3}a_{4}}^{+}-\lambda _{a_{1}a_{2}a_{3}a_{4}}^{0},
\label{4pt_alternative}
\end{eqnarray}%
with the constraint
\begin{equation}
  \widetilde{\lambda }_{a_{1}a_{2},a_{3}a_{4}}^{+}+\widetilde{\lambda }_{a_{1}a_{3},a_{2}a_{4}}^{+}+\widetilde{\lambda }_{a_{2}a_{3},a_{1}a_{4}}^{+}=0.  
\end{equation}
Similar decomposition can be written also for 5pt amplitudes.
Therefore in the case of the 4pt and 5pt amplitudes we can distinguish two types of contact terms. 
The first one is completely symmetric with respect to the permutation of the momenta and is represented by the Gram determinant (cf. (\ref{vertex})). 
There is always one such monomial for every combination of flavors and depending on the particular combination of flavors there are eventually other terms.
The latter we call of the second type. 
This decomposition is not unique since it depends on the actual choice of basis of the second-type contact terms. 
For instance, for the 4pt amplitude $A_4(1^a,2^a,3^b,4^b)$ we can use either the decomposition (\ref{4pt_decomposition}), where explicitly
\begin{equation}
 A_4(1^a,2^a,3^b,4^b)=(\lambda_{aa,bb}^{+}+2\lambda_{ab,ab}^{+})s_{12}s_{13}s_{23}+ \frac{1}{3}(\lambda_{aa,bb}^{+}-\lambda_{ab,ab}^{+}) (2s_{12}^{3}-s_{13}^{3}-s_{23}^{3}),
\end{equation}
or the alternative  ones, namely
\begin{equation}
 A_4(1^a,2^a,3^b,4^b)=3\lambda_{aa,bb}^{+} s_{12}s_{13}s_{23}+(\lambda_{ab,ab}^{+}-\lambda_{aa,bb}^{+})( s_{13}^{3}+s_{23}^{3}),
\end{equation}
and (cf. (\ref{2_flavor_4pt}))
\begin{equation}
 A_4(1^a,2^a,3^b,4^b)=3\lambda_{ab,ab}^{+} s_{12}s_{13}s_{23}+(\lambda_{aa,bb}^{+}-\lambda_{ab,ab}^{+}) s_{12}^{3}.
 \label{4pt_modified_new}
\end{equation}
Nevertheless, provided the amplitudes correspond to those derived from the Lagrangian (\ref{Lagrangian_basic}) without cubic interactions, the second-type contact terms vanish identically irrespectively on the choice of the basis.
Let us note, that for our numerical implementation of the bonus relation, the actual choice of the basis of the second-type terms, for which we use an appropriate optimized algorithm, is not essential\footnote{The flavor content of the 4pt and 5pt amplitudes as well as  the number of the independent second-type terms as a function of the number of flavors is discussed in detail in Appendix \ref{number_2nd_type_terms}. }. 

Using this terminology, the results of the analysis of the two-flavor case performed in Section \ref{applications_section} can be now reformulated as follows:
The necessary condition for the $O(p^2)$ behavior of the amplitudes of the two-flavor theory with Galileon power counting  is the absence of second-type terms in all the 4pt amplitudes.
It is a natural question whether the analogous criterion is valid also for the 5pt amplitudes and in the general $N$-flavor case.
We, therefore, tested the following conjecture numerically:

\emph{In the general $N$-flavor theories with Galileon power counting in $D=4$ dimensions, the necessary condition for the enhanced $O(p^2)$ Adler zero is the absence of the contact terms of the second type in all the 4pt and 5pt amplitudes.}

The validity of the first part of this conjecture concerning the 4pt amplitudes was tested using the bonus relations for 6pt amplitudes as explained in the introduction to this section.
We have employed an algorithm with an automatic generation of the basis of the 4pt second-type terms as well as numerical kinematics and probing the corresponding bonus relations we found evidence for the validity of the conjecture up to and including five flavors\footnote{Note that we need at least four different flavours to capture the term with totally antisymmetric coupling in the 4pt amplitude (\ref{4pt_alternative})}.

The second part of the above conjecture concerning the 5pt amplitudes has been tested  for $N=5$ by means of an automatic creation  of the basis of the second-type 5pt contact terms and checking the $O(p^2)$
enhanced soft limit for the complete 5pt amplitude with all the possible flavor contents.
Details of this algorithm in the case of the direct check of soft behavior for the 5pt vertices can be found in \cite{Cheung:2016drk}. 
Again we can summarize that in all tested cases we have been in agreement with the conjecture\footnote{Similar result for two flavors was obtained in \cite{Elvang:2018dco}.}.
Note that the case $N=5$ covers all the possible patterns of the Bose symmetry which can be associated with  different flavor structures of the 5pt amplitudes.
Therefore, since the 5pt amplitudes are contact, probing all the five-flavor 5pt amplitudes for the soft behavior is in fact a test of the general $N$ case. 
We can therefore conclude that the second part of the conjecture  has been proven numerically.
This is, however, not the case of the first part of the conjecture, the probe of which uses the 6pt amplitudes. 
Note that the latter  have  pole contributions corresponding to the one-particle exchanges and these depend on the actual number of flavors.
Threfore, though we have strong evidence for the validity of the first part of the conjecture, we cannot consider it definitely proven. 

Note, however, that if true, the conjecture gives not only the necessary, but also sufficient conditions.
Indeed, provided these conditions are satisfied, the 4pt and 5pt amplitudes are those of the multi-flavor Galileon theory (\ref{Lagrangian_basic}). They obey the enhanced $O(p^2)$ Adler zero, and using the soft BCFW relations, which implement the enhanced soft behavior, we can construct all the remaining tree-level amplitudes unambiguously.

\subsection{Uniqueness of the $U(N)$ Galileon}
In the construction of the  monomials entering the seed amplitudes we had to enforce a full permutation symmetry within the subclasses of the same flavors. 
As we know from explicit $U(N)$ example introduced in Sec.~\ref{sec:explicitexamples}, it is possible to considerably simplify calculations of flavor amplitudes for some models using the stripped amplitudes. 
These stripped amplitudes are  cyclically ordered objects in external momenta. 
We can as well focus on them from the bottom-up perspective of this section and test whether these stripped amplitudes are fixed uniquely by the soft theorem of the form (\ref{eq:softstrip}).

\begin{figure}[t]
  \centering
    \includegraphics[scale=0.66]{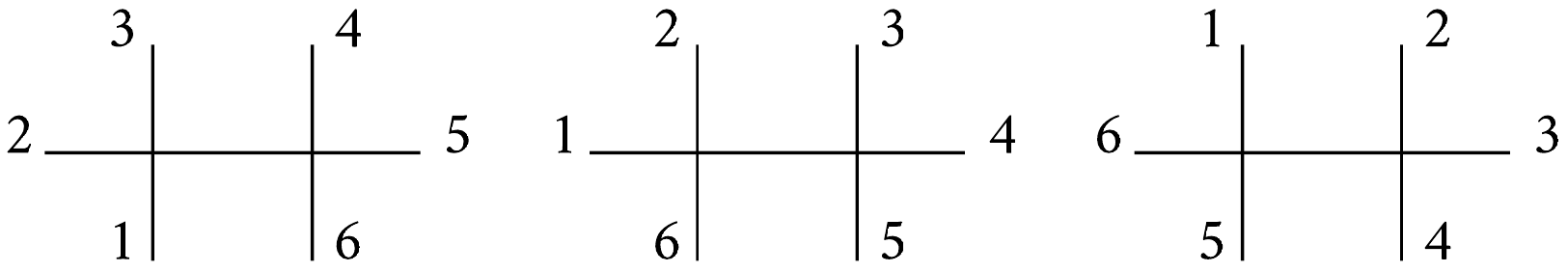}
    \caption{The factorization diagrams for the ordered 6pt amplitude}
    \label{fig:fac6}
\end{figure}

For this purpose we have first constructed the  basis of the cyclically ordered 4pt, 5pt and 6pt vertices. 
Their explicit form is again not important, but we have to make sure that the basis is complete. 
For reference, working in $D=4$ dimensions,  we obtain the following number of terms: 2, 14 and 225 for the 4pt, 5pt and 6pt vertices, respectively. 
Using these vertices, the most general ordered 6pt amplitude $\mathcal{A}_6(1,2,3,4,5,6)$ can be then easily calculated, apart from the contact terms this includes the factorization graphs depicted in Fig.~\ref{fig:fac6}.
This amplitude depends then on 241 free coupling constants.

With this general basis in hand we can revisit the $U(N)$ Galileon defined by Lagrangian~(\ref{SU(N)_Galileon_Lagrangian}). 
We have learned that it is possible to obtain the corresponding amplitudes from the stripped objects that have cyclic symmetry. 
For these objects, the stripped amplitudes defined by  (\ref{stripped_amplitudes_definition}), we have proved the soft theorem with a non-trivial right-hand side~(\ref{eq:softstrip}) with free $\lambda_3$. 
We can now turn it upside down and ask the following important question: how many theories comply with this soft theorem. 
From the soft-bootstrap consideration it is clear that for our particular Galileon power counting any amplitude can be reconstructed using the soft BCFW recursion relations out of the 4pt and 5pt amplitudes only. 
For the most general 6pt amplitude constructed above it thus means that all 225  constants mentioned above and corresponding to the contact 6pt vertices must be fixed by imposing the soft theorem of the form (\ref{eq:softstrip}). 
We have explicitly numerically verified this. 
Along this process we have also found additional relations for the 4pt and 5pt constants that must be fulfilled. 
These self-consistency relations pick a unique solution with only two additional parameters on the top of $\lambda_3$ which enters the right-hand side of the soft theorem (\ref{eq:softstrip}).
These two additional constants have thus to be  in unique correspondence with the $U(N)$  Galileon couplings $\lambda_4$ and $\lambda_5$. 
We have thus proved a uniqueness of this theory. 
Trivially if the right-hand side of the soft theorem~(\ref{eq:softstrip}) is zero we have again a unique theory which is equivalently to~(\ref{SU(N)_Galileon_Lagrangian}) with $\lambda_3=0$ (no three-point vertices).

\section{Summary and conclusions}\label{sec:summary}
In this paper we have discussed two possible definitions of the multi-flavor Galileon theories.
The first one was based on the Lagrangian (\ref{Lagrangian_basic}), from which we derived the basic properties of the tree-level scattering amplitudes, namely the  nontrivial soft theorem (\ref{soft_theorem_galileon}) for the single-particle soft limit. 
The latter turned out to be a generalization of the analogous soft theorem \cite{Kampf:2019mcd}, originally found for the Goldstone bosons of the nonlinear sigma models with power counting $\rho=0$, to the case of the theories with power counting $\rho=2$.
The second possible definition we have discussed is the amplitude-oriented one which is  based on the fixed power counting and the single-particle soft limit represented by the above mentioned soft theorem.

Our soft theorems were demonstrated on two explicit examples of multi-flavor Galileon theories. 
The first example, the $U(N)$ Galileon, is already known theory where it is possible to ``strip'' the group structure  using the cyclically ordered amplitudes. 
We have shown that also for these stripped amplitudes  we can  get a closed form of the soft theorem. 
We have also proved using the amplitude methods that this soft theorem is a defining property that fixes unambiguously the theory.
The second example represents a new three-flavor Galileon theory, which, to our knowledge, has not yet been discussed in the literature. Based on the construction we have dubbed it $U(2)/U(1)$ Galileon. 
It is an example of the multi-Galileon theory with one neutral and two charged fields, which mixes both the trivial soft theorem (for the neutral particle) and the generalized one (for the charged particles). It represents an analogue of the known example  of the fibrated $CP^1$ sigma model, now for the Galileon counting. 

We have also discussed in detail under which circumstances the soft theorem (\ref{soft_theorem_galileon}) trivializes yielding $O(p^2)$ behavior of the amplitudes in the appropriate basis of the one-particle states.
Such a criterion is tightly connected with the existence of a duality transformation that removes the cubic vertices from the Lagrangian without changing the form of the remaining interactions up to a change of the coupling constants and up to a total derivative.
We have found that such a duality exists only when the Lagrangian, after an appropriate $SO(N)$ rotation of the fields, becomes a sum of the Lagrangians of two noninteracting sectors, namely  of the Lagrangian of the form (\ref{Lagrangian_basic}) without cubic vertices and of a sum of several decoupled single-Galileon ones with possible cubic vertices.

Then we formulated  appropriately  modified soft BCFW recursion relations based on the soft theorem (\ref{soft_theorem_galileon}). 
These can be used for the reconstruction of all the tree-level amplitudes from the seed amplitudes (i.e. from the known 4pt and 5pt amplitudes in $D=4$ dimensions, or 4pt,\,$\dots$, $(D+1)$pt ones in the general case).
As a consequence, the multi-Galileon theories based on the Lagrangian (\ref{Lagrangian_basic}) can be defined alternatively as a \emph{particular solution} of the corresponding soft bootstrap, i.e. of the problem of bottom-up amplitude-based construction of the theories with the Galileon power-counting and satisfying the soft theorem (\ref{soft_theorem_galileon}).
Note, however, that the seed amplitudes for such construction are in this particular case derived from the Lagrangian (\ref{Lagrangian_basic}) and therefore they are not the most general ones.

Therefore,  it is not possible to exclude \emph{a priori} the existence of other solutions of the above problem, which are not equivalent to the multi-Galileon theories defined by the Lagrangian (\ref{Lagrangian_basic}).
If this were the case, the two above definitions of the multi-Galileon theories, namely the Lagrangian-based and the amplitude-based, would not be equivalent.

We, therefore, tested the uniqueness of such a solution  allowing for the most general seed amplitudes with the Galileon power counting as a starting point of the recursion and probing the self-consistency of the corresponding bottom-up recursive construction.
For this purpose we have used a method based on bonus relations. These express the necessary conditions for the existence of the healthy  higher-order amplitudes obtained by the recursion from the seed ones. 
As a consequence, the bonus relations impose nontrivial constraints on the seed amplitudes which can be solved, in some special cases even analytically.

For simplicity we restricted ourselves to the case of the trivial soft theorem, i.e. to the requirement of the $O(p^2)$ soft behaviour of the amplitudes.
In the two-flavor case, we were able to solve the constraints implied by the bonus relation for 6pt amplitudes  analytically and proved that the only solution for the 4pt seed amplitudes is that they originate from the Lagrangian  (\ref{Lagrangian_basic}) without the cubic terms.
This result was a basis for the conjecture that the same is true also for a general number of flavors, namely that for the 4pt seed amplitudes all the contact terms of the second type (i.e. those which can not be produced by the Lagrangian (\ref{Lagrangian_basic}) without cubic terms) must vanish in order to satisfy the 6pt  bonus relations. 
This conjecture has been tested numerically in $D=4$ dimensions  and up to and including five flavors we have found evidence of its validity.
Similarly we have tested and numerically proved analogous conjecture concerning the 5pt amplitudes, namely that the absence of the second-type terms is a necessary condition for the $O(p^2)$ behavior of  the 5pt amplitudes in the single-particle soft limit.

We have also discussed the possibility of the existence of the multi-flavor version of the Special Galileon with $O(p^3)$ soft limit of the amplitudes.
In $D=4$ we tried to found it as a special case of the theory with 4pt seed amplitudes without second-type terms, since otherwise, according to our conjecture, already the $O(p^2)$ behavior would not apply.
The solution of the consistecy relation for the 6pt amplitudes has been found analytically with the result that, up to a $SO(N)$ rotation of the fields, the only possibility is a sum of decoupled single-flavor Special Galileons.
Provided our conjecture concerning the criterion of the $O(p^2)$ behavior is true for general $N$, there is then no nontrivial multi-flavor version of the Special Galileon.

\appendix
\section{Kinematical factors in the two-flavor bonus relations}\label{kinematical factors}

In this appendix we give  explicit formulas for the kinematical factors entering the bonus relations of the two-flavor case.
For the relation (\ref{bonus_relation_2}) we get
\begin{eqnarray}
K^{\left( I\right) }\left( s_{ij}\right) &=&\frac{1}{3!}\sum_{\sigma \in
S_{4}}3s_{12}s_{1\sigma(3)}s_{2\sigma(3)}s_{\sigma(4)\sigma(5)}s_{\sigma(4)\sigma(6)}s_{\sigma(5)\sigma(6)} 
\notag\\&&\times\left[ \frac{1}{s_{2\sigma(3)}\left(
a_{\sigma(3)}-a_{2}\right) }+\frac{1}{s_{1\sigma(3)}\left( a_{\sigma(3)}-a_{1}\right) }\right],
\label{kin1}\\
K^{\left( II\right) }\left( s_{ij}\right) &=&\frac{1}{4}\sum_{\sigma \in
S_{4}}3s_{1\sigma \left( 3\right) }s_{1\sigma \left( 4\right) }s_{\sigma
\left( 3\right) \sigma \left( 4\right) }s_{2\sigma \left( 5\right)
}s_{2\sigma \left( 6\right) }s_{\sigma \left( 5\right) \sigma \left(
6\right) }  \notag \\
&&\times \left[ \frac{1}{s_{1\sigma \left( 3\right) }\left( a_{1}-a_{\sigma
\left( 3\right) }\right) }+\frac{1}{s_{1\sigma \left( 4\right) }\left(
a_{1}-a_{\sigma \left( 4\right) }\right) }\right.  \notag \\
&&\left. +\frac{1}{s_{2\sigma \left( 5\right) }\left( a_{2}-a_{\sigma \left(
5\right) }\right) }+\frac{1}{s_{2\sigma \left( 6\right) }\left(
a_{2}-a_{\sigma \left( 6\right) }\right) }\right] , \label{kin2}\\
K^{\left( III\right) }\left( s_{ij}\right) &=&\frac{1}{4}\sum_{\sigma \in
S_{4}}\frac{s_{\sigma \left( 3\right) \sigma \left( 4\right) }s_{\sigma
\left( 5\right) \sigma \left( 6\right) }}{\left( a_{1}-a_{2}\right) }  \notag
\\
&&\times \left\{ s_{\sigma \left( 5\right) \sigma \left( 6\right)
}s_{1\sigma \left( 3\right) }s_{1\sigma \left( 4\right) }\left[ s_{2\sigma
\left( 5\right) }\left( a_{1}-a_{\sigma \left( 5\right) }\right) +s_{2\sigma
\left( 6\right) }\left( a_{1}-a_{\sigma \left( 6\right) }\right) \right] %
\phantom{\frac{1}{2}}\right.  \notag \\
&&\left. \times \left[ \frac{1}{s_{1\sigma \left( 4\right) }\left(
a_{1}-a_{\sigma \left( 4\right) }\right) }+\frac{1}{s_{1\sigma \left(
3\right) }\left( a_{1}-a_{\sigma \left( 3\right) }\right) }\right] \right. 
\notag \\
&&\left. -s_{\sigma \left( 3\right) \sigma \left( 4\right) }s_{2\sigma
\left( 5\right) }s_{2\sigma \left( 6\right) }\left[ s_{1\sigma \left(
3\right) }\left( a_{2}-a_{\sigma \left( 3\right) }\right) +s_{1\sigma \left(
4\right) }\left( a_{2}-a_{\left( 4\right) }\right) \right] \right.  \notag \\
&&\left. \times \left[ \frac{1}{s_{2\sigma \left( 5\right) }\left(
a_{2}-a_{\sigma \left( 5\right) }\right) }+\frac{1}{s_{2\sigma \left(
6\right) }\left( a_{2}-a_{\sigma \left( 6\right) }\right) }\right] \right\}.
\label{kin3}
\end{eqnarray}%
In the case of the bonus relation (\ref{bonus_relation_3}) we have
\begin{eqnarray}
K^{\left( IV\right) }\left( s_{ij}\right) &=&3\sum_{\sigma ,\rho \in
S_{3}}s_{\sigma \left( 1\right) \sigma \left( 2\right) }s_{\sigma \left(
1\right) \rho \left( 4\right) }s_{\sigma \left( 2\right) \rho \left(
4\right) }s_{\sigma \left( 3\right) \rho \left( 5\right) }s_{\sigma \left(
3\right) \rho \left( 6\right) }s_{\rho \left( 5\right) \rho \left( 6\right) }
\notag \\
&&\times \left[ \frac{1}{s_{\sigma \left( 3\right) \rho \left( 5\right)
}\left( a_{\sigma \left( 3\right) }-a_{\rho \left( 5\right) }\right) }+\frac{%
1}{s_{\sigma \left( 3\right) \rho \left( 6\right) }\left( a_{\sigma \left(
3\right) }-a_{\rho \left( 6\right) }\right) }\right],  \notag \\
K^{\left( V\right) }\left( s_{ij}\right) &=&3\sum_{\sigma ,\rho \in
S_{3}}s_{\sigma \left( 1\right) \sigma \left( 2\right) }s_{\sigma \left(
1\right) \rho \left( 4\right) }s_{\sigma \left( 2\right) \rho \left(
4\right) }s_{\sigma \left( 3\right) \rho \left( 5\right) }s_{\sigma \left(
3\right) \rho \left( 6\right) }s_{\rho \left( 5\right) \rho \left( 6\right) }
\notag \\
&&\times \left[ \frac{1}{s_{\sigma \left( 1\right) \rho \left( 4\right)
}\left( a_{\rho \left( 4\right) }-a_{\sigma \left( 1\right) }\right) }+\frac{%
1}{s_{\sigma \left( 2\right) \rho \left( 4\right) }\left( a_{\rho \left(
4\right) }-a_{\sigma \left( 2\right) }\right) }\right].  \label{kin4}
\end{eqnarray}%

\section{Combinatorics of the 4pt and 5pt vertices}\label{number_2nd_type_terms}

Let us first focus on the 4pt vertices. Their form can be read-off e.g.  from~(\ref{4pt_decomposition}). 
For the same-flavor combination, i.e. $aaaa$ there is only one possibility, namely $G(1,2,3)\sim s_{12}s_{13}s_{23}$, and similarly for the same-but-one ($aaab$) combination. 
For the $aabb$ combination there are in total two terms, one of the second type. 
This can be chosen e.g. in the  form $\sim s_{12}^3$. 
And similarly we can proceed for other combinations.
In total the number of monomials for all possible individual combinations of different flavors can be summarized as
\begin{equation}
aaaa:1,\; aaab:1,\; aabb:2,\; aabc:2,\; abcd:4\,. 
\label{eq:no4ptfl}
\end{equation}
Number of different 4pt amplitudes is given by the number of combinations with repetition for $N$ flavors
\begin{equation}
K(4,N) = \binom{N+3}{4} = \frac{1}{4!} N(N+1)(N+2)(N+3)\,.    
\end{equation}
We can check that for $N=1$ this is 1, corresponding to $aaaa$. For $N=2$ it is 5 corresponding to $aaaa$, $aaab$, $aabb$, $abbb$, $bbbb$ and so on for higher $N$.
As there is always one symmetric vertex for each combination, this number also represents a number of the first-type constants.

Including all possible combinations of~(\ref{eq:no4ptfl}) the total number of 4pt vertices is given\footnote{Note that this formula represents partially summed centered tetrahedral numbers.}
\begin{equation}
    \# \text{4pt vertices }= \frac{1}{6}N^2 (N^2+5).
\end{equation}
Subtracting the number of the first-type vertices, we get number of extra terms:
\begin{equation}
    \# \text{of second type} = \frac{1}{8}N (N-1)(N^2-N+2)\,.
\end{equation}
Similarly the situation with the 5th vertices can be summarized as follows
\begin{equation}
aaaaa:2,\; aaaab:5,\; aaabb:9,\; aaabc:14,\; aabbc:22,\; aabcd:28,\; abcde:70\,. 
\end{equation}
Number of different 5pt vertices (i.e. constants of the first type) is given by the number of combinations with repetition for $N$ flavors
\begin{equation}
    K(5,N) = \binom{N+4}{5} = \frac{1}{5!} N(N+1)(N+2)(N+3)(N+4)\,.
\end{equation}
Including all possible combinations, the total number of 5pt vertices is given
\begin{equation}
    \# \text{5pt vertices }= \frac{1}{12} N (7N^4 - 14 N^3 + 125N^2 - 214 N +120)\,.
\end{equation}
Subtracting the number of the first type, we get number of extra terms:
\begin{equation}
   \# \text{of second type} = \frac{1}{40}N (23 N^4-50 N^3+405 N^2-730 N+392)\,.
\end{equation}   

\section{Bonus relations for the multi-flavor Special Galileon}\label{appendix_bonus_relations}
In this Appendix we derive the bonus relations, which we mentioned in Subsection \ref{sec:explicitexamples}.
Assume that the 6pt amplitudes have $O\left( p^{3}\right) $ single soft limit.
Then the function
\begin{equation}
f(z) =\frac{A_{6}\left( p_{1}^{a_{1}}(z) ,\ldots
,p_{6}^{a_{6}}(z) \right) }{\prod\limits_{j=1}^{6}\left(
1-a_{j}z\right) ^{3}}\overset{z\rightarrow \infty }{=}O\left( z^{-8}\right)
\end{equation}
has only the unitarity poles.
Since it behaves better than $1/z$ for $z\to\infty$,  we have bonus relations
\begin{equation}
\sum_{\mathcal{F},I=\pm }\mathrm{res}\left( z^{k}f(z) ,z_{\mathcal{F}}^{I}\right) =0\,,
\end{equation}
for $k=0,1,2,\ldots ,6$. 
Using the residue theorem, let us rewrite them in the form 
\begin{equation}
\sum_{\mathcal{F},I=\pm }\mathrm{res}\left( z^{k}f(z) ,z_{%
\mathcal{F}}^{I}\right) =\sum_{\mathcal{F}}\sum_{l=1}^{6}\mathrm{res}\left(
z^{k}f_{\mathcal{F}}(z) ,\frac{1}{a_{l}}\right) =0\,,
\end{equation}
where for given factorization channel, e.g. $\mathcal{F=}\left\{1^{a_{1}},2^{a_{2}},3^{a_{3}}\right\}$
\begin{eqnarray}
f_{\mathcal{F}}(z) &=&\lambda _{a_{1}a_{2}a_{3}a}^{+}\lambda
_{a_{4}a_{5}a_{6}a}^{+}\frac{s_{12}(z) s_{13}(z)
s_{23}(z) s_{45}(z) s_{46}(z)
s_{56}(z) }{p_{\mathcal{F}}^{2}(z)
\prod\limits_{j=1}^{6}\left( 1-a_{j}z\right) ^{3}} \notag\\
&=&\lambda _{a_{1}a_{2}a_{3}a}^{+}\lambda _{a_{4}a_{5}a_{6}a}^{+}\frac{%
s_{12}s_{13}s_{23}s_{45}s_{46}s_{56}}{p_{\mathcal{F}}^{2}(z)
\prod\limits_{j=1}^{6}\left( 1-a_{j}z\right) }\,.
\end{eqnarray}
The bonus relations read then up to an overall combinatoric factor
\begin{eqnarray}
 &&\sum_{\sigma \in S_{6}}\lambda _{a_{\sigma \left( 1\right) }a_{\sigma
\left( 2\right) }a_{\sigma \left( 3\right) }a}^{+}\lambda _{a_{\sigma \left(
4\right) }a_{\sigma \left( 5\right) }a_{\sigma \left( 6\right) }a}^{+}{s_{\sigma \left( 1\right) \sigma \left( 2\right) }s_{\sigma
\left( 1\right) \sigma \left( 3\right) }s_{\sigma \left( 4\right) \sigma \left( 5\right) }s_{\sigma
\left( 4\right) \sigma \left( 6\right) }s_{\sigma \left( 5\right) \sigma
\left( 6\right) }}
  \notag \\
&&\times \frac{a_{\sigma \left( 1\right) }^{8-k}}{\left( a_{\sigma \left( 1\right) }-a_{\sigma \left( 2\right)
}\right) \left( a_{\sigma \left( 1\right) }-a_{\sigma \left( 3\right)}\right)\prod\limits_{j\neq \sigma \left( 1\right) }\left(
a_{\sigma \left( 1\right) }-a_{j}\right) }=0\,.  
\end{eqnarray}
Comparing this with the explicit form of the amplitude (\ref{amplitude_special_multi_galileon}) (see also (\ref{pole_term}) and (\ref{counterterm})), we find that these relations are equivalent to the requirement of the independence of the amplitude (\ref{amplitude_special_multi_galileon}) on the shifts $a_j\to a_j+a$.
\acknowledgments
We acknowledge valuable collaboration with Mikhail Shifman and Jaroslav Trnka on some general aspects connected with this project.
This work is supported in part by the Czech Government projects GACR 18-17224S and LTAUSA17069.

\providecommand{\href}[2]{#2}\begingroup\raggedright
\endgroup

\end{document}